\documentclass[letter]{aa} 
\usepackage{placeins}
\usepackage{tabularx} 
\usepackage{booktabs}  

\usepackage{mdframed}
\usepackage{adjustbox}
\usepackage{array}  
\usepackage{float} 
\usepackage{graphicx}
\usepackage{txfonts}
\usepackage{natbib}
\usepackage{soul}
\usepackage{float}
\usepackage{graphicx}	
\usepackage{amsmath}	
\usepackage{amssymb}
\usepackage[T1]{fontenc}
\usepackage{subcaption}
\usepackage{caption}

\DeclareUnicodeCharacter{2212}{-}
\DeclareUnicodeCharacter{2032}{-}
\usepackage{longtable}
\DeclareRobustCommand{\VAN}[3]{#2}
\let\VANthebibliography\thebibliography
\def\thebibliography{\DeclareRobustCommand{\VAN}[3]{##3}\VANthebibliography}
\begin{document} 
\title{AstroSat/UVIT Study of NGC~663: First detection of Be+sdOB systems in a young star cluster}

  \subtitle{}

   \titlerunning{}

\author{
Sneha Nedhath\inst{1,2,3}\thanks{e-mail: snehanedhath@gmail.com}
\and
Sharmila Rani\inst{1,4}
\and
Annapurni Subramaniam\inst{3}
\and
Elena Pancino\inst{1}
}

\institute{
INAF — Osservatorio Astrofisico di Arcetri, Largo E. Fermi 5, I-50125 Firenze, Italy
\and
Dipartimento di Fisica e Astronomia, Università di Firenze, Via G. Sansone 1, I-50019 Sesto Fiorentino (Firenze), Italy
\and
Indian Institute of Astrophysics, Koramangala, Bangalore 560034, India
\and
Aryabhatta Research Institute of Observational Sciences (ARIES), Manora Peak, Nainital 263002, India
}

\authorrunning{Nedhath et al.}
\date{}

% \abstract{}{}{}{}{} 
% 5 {} token are mandatory
 
\abstract
{Be stars are rapidly rotating stars surrounded by a disc; however, the origin of these stars remains unclear. Mass and angular momentum transfer in close binaries account for the rapid rotation of a major fraction of Be stars, as supported by the previous detection of low-mass stripped companions to these stars. The stripped companions can be helium-burning subdwarf OB-type stars (sdOBs) and white dwarfs.}
{ The main objective of this study is to characterise the identified Be stars in the young open cluster NGC~663 and search for possible hot companions.}
{We present the first ultraviolet (UV) photometric study of NGC~663 using far-UV and near-UV data from UVIT/\textit{AstroSat} as a part of the UOCS series (XVIII). We identified 23 previously known Be stars in the cluster. Further, we utilised the spectral energy distribution fitting technique to derive the fundamental parameters and to search for UV-bright companions of the  identified Be stars.}
{Our study reveals that 19 out of 23 Be stars show a significant UV excess, indicating the presence of hot companions. Here, we report the first detection of high-mass sdOB companions to Be stars, with 69.5\% of them found in binaries within a cluster, offering direct evidence of binary interactions. }
{ 
This study showcases the key role of binary interactions in the formation of Be stars in clusters and provides insights into massive star evolution.
}


\maketitle

\section{Introduction}
Classical B emission-line stars (Be stars, hereafter) are rapidly rotating main-sequence (MS) B-type stars surrounded by ionised, gaseous discs. 
The exact mechanism behind the formation of these discs, commonly known as the 'Be phenomenon', remains elusive. Various mechanisms have been proposed to explain how the material is expelled from the Be star into the disc, including rotation, non-radial pulsation, magnetic fields, and the presence of binary companions \citep{2013A&ARv..21...69R,2020A&A...641A..42B,2008MNRAS.388.1879M}.\\ 

The binary formation pathway suggests the crucial role of a companion star that fills its Roche lobe, leading to the formation of a circumstellar disc around the Be star. Based on simulations of binary star populations, \cite{2013ApJ...764..166D} suggested that rapidly rotating massive stars could be spun up either through mass transfer or mergers. Additionally, \cite{2014ApJ...782....7D} concluded that approximately 30\% of all MS B-type stars are products of dynamical interactions occurring among the stars in clusters. Binary population synthesis studies specific to Be stars \citep{1991A&A...241..419P,1997A&A...322..116V,2014ApJ...796...37S,2021ApJ...908...67S} also suggest that at least a significant fraction, if not all, are likely the products of binary interactions. \\ 
A series of studies conducted by \cite{2008ApJ...686.1280P, 2013ApJ...765....2P, 2016ApJ...828...47P} also support the existence of invisible companions that play a role in the formation of discs in Be stars. Hot companions of these stars are faint in optical and infra-red (IR) but are easily detectable in the far-ultraviolet (FUV) as they emit most of their flux at short wavelengths. Recently, \cite{2021AJ....161..248W} increased the sample of known Be+sdO binaries.

The majority of the Be-companions are hot subluminous stars of subdwarf O, B-type (sdOB) stars \citep{2022ApJ...940...86K}. These companions can subsequently evolve into white dwarfs (WDs) or neutron stars (NSs), potentially leading to the formation of Be/X-ray binaries (BeXRBs). Some of these post-mass transfer (post-MT) systems have already been identified as Be-NS or Be-WD systems through X-ray outbursts. To detect stripped stars next to an optically bright companion, \cite{2018A&A...615A..78G} recommended systematic searches for their UV excess. This clearly showcases the importance and need to employ UV imaging to shed more light on their binary formation pathway. So far, no Be+sdOB system has been detected in star clusters.\\

Young open clusters (OCs) with an age of less than a hundred million years are ideal laboratories for the study of Be stars among the B-type stars and the role of binarity in shaping their properties. In the study by \cite{2008MNRAS.388.1879M} covering emission line stars in 42 young OCs, NGC~663 stood out as one of the clusters with a large number of Be stars. NGC~663 is a young OC located in the constellation Cassiopeia. \cite{2001A&A...376..144P} estimated an age of 20-25 Myr and a distance of 2.1 kpc. Recently, \cite{2021MNRAS.504..356D} estimated the cluster distance to be 2.353 kpc, log(t) = 7.403 yr, \big[Fe/H\big] = -0.125\,dex, and A$_{V}$ = 2.268\,mag, which are adopted in this work.\\
This study investigates the ultraviolet properties of Be stars in NGC~663 to check for the presence of hot companions and reports the first detection of binary Be stars with subdwarf companions in OCs, providing evidence for such a formation pathway for Be phenomena in OCs.

\section{Data reduction and analysis}
\label{data}
We used archival data of UVIT onboard \textit{AstroSat} (L1 data from the \textit{AstroSat} data archive\footnote{\url{https://astrobrowse.issdc.gov.in/astro_archive/archive/Home.jsp}}) and processed them using  CCDLAB \cite{2017PASP..129k5002P}. 
A detailed description of UVIT and its calibration can be found in \cite{2017AJ....154..128T, 2020AJ....159..158T}. NGC~663 was observed with UVIT on 12 August, 2017, (Proposal Id: A03-079) in three FUV (F148W, F154W, F169M) and two NUV filters (N263M and N279N).  
The CCDLAB corrects for satellite drift, flat field, distortion, fixed pattern noise, and cosmic rays. The orbit-wise images are then co-aligned and merged to get the final science-ready images. Details of the UVIT observations of NGC~663 are tabulated in Table~\ref{uvit_combined}. Figure~\ref{uv_color_image} shows a composite two-colour image of NGC~663 (F148W and N279N).

% \FloatBarrier
\section{Photometry}
\label{phot}
Aperture and point spread function (PSF) photometry were carried out using the DAOPHOT package in IRAF/NOAO \cite{1987PASP...99..191S}. The model PSF was constructed using bright, isolated stars and then applied to all the detected stars using the ALLSTAR task to obtain PSF-ﬁtted magnitudes. The PSF, aperture, and saturation corrections were subsequently applied to obtain the instrumental magnitudes. We note that many bright stars in the cluster were too saturated to derive reliable magnitudes.

The details of the saturation correction are described in \citep{2017AJ....154..128T}. Using zero-point (ZP) magnitudes reported in the calibration paper \citep{2020AJ....159..158T}, the derived instrumental magnitudes were calibrated into the AB magnitude system. By limiting the PSF fitted error to $\sim$0.2\,mag (see Fig.~\ref{psf_fit_err}), stars detected down to $\sim$21\,mag in the F148W, F154W, N236M filters, and down to $\sim$19\,mag in the F169M and N279N filters, were selected for further analysis.
We used the reddening value of E(B$-$V) = 0.7\,mag \citep{2021MNRAS.504..356D} and  R$_{V}$=3.1 from \cite{1958AJ.....63..201W} for the Galaxy. Given the differential extinction within the cluster reported by \cite{2001MNRAS.328..370Y}, we used a range in the A$_{v}$ values with the corresponding extinction in the respective filters tabulated in Table~\ref{uvit_combined} to overlay the isochrones.

\begin{figure}
    \centering
    \includegraphics[scale=0.41]{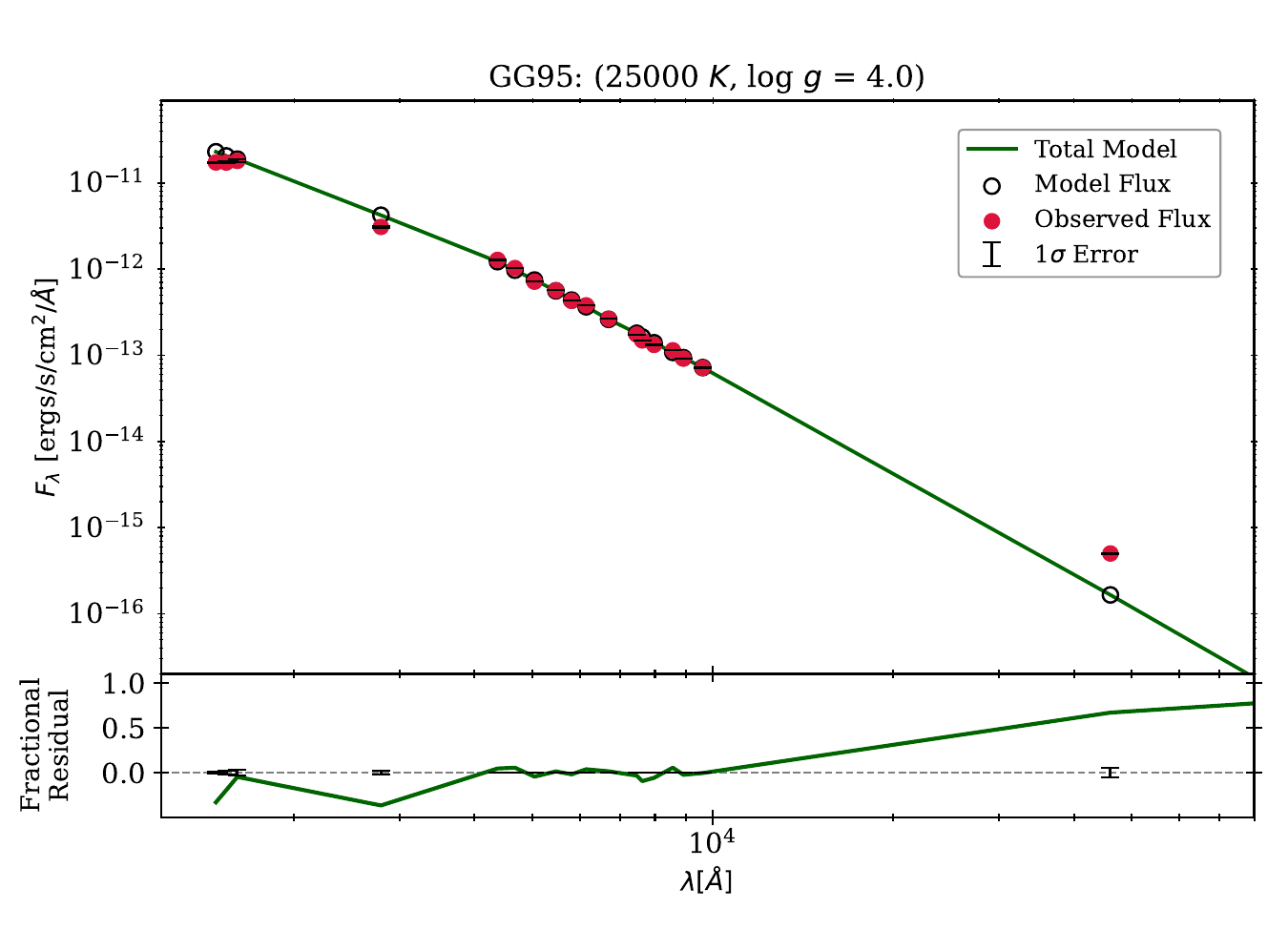}
    \includegraphics[scale=0.41]{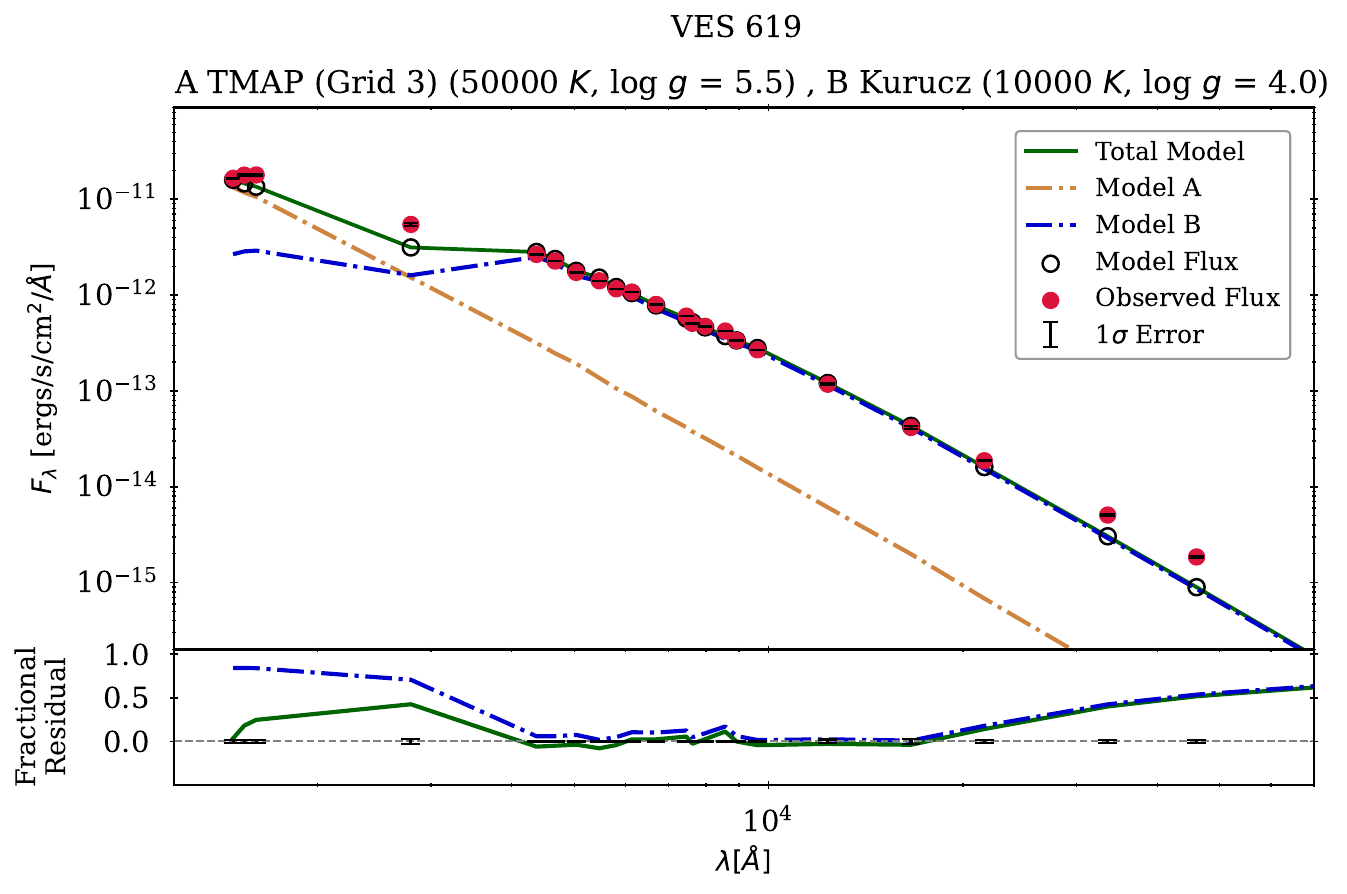}
    \caption{Upper panel: Single-fit SED of Be star GG 95, with the Kurucz model (green line) fitted to observed fluxes (red). Lower panel: SED of Be star VES 619, fitted with a combined Kurucz (dash-dotted blue line, cool) and TMAP (dash-dotted brown line, hot) model. The composite fit here is shown with a solid green line. Best-fit parameters and model details are shown above each plot.}
    \label{sed fit}
\end{figure}

\section{Be stars in NGC~663}
\label{cluster}

\cite{1979AJ.....84.1319S} found 27 H$\alpha$ emission stars, mostly earlier than the B5 spectral type. This number was later updated to 24, with 12 stars showing variability in H$\alpha$ emission \citep{1990AJ....100.1239S}. 
Subsequently, \cite{2001yCat..33760144P} identified 26 Be stars, the majority of which are between B0 - B3. Later, using the slitless spectroscopy method, \cite{2008MNRAS.388.1879M} detected several new Be stars, increasing the total count in the cluster to 31, within the B0-B8 range.  
Among the 31 known Be stars, \cite{2015AJ....149...43Y} re-identified 15 stars and discovered four new Be stars. They concluded that NGC~663 contains a total of 34 Be stars. Of these, 23 stars were detected by UVIT which are analysed in this study.

\section{Spectral energy distributions of Be stars}
\label{sed}
To derive the stellar parameters of the Be stars detected with UVIT, we constructed their spectral energy distributions (SEDs) using the Virtual Observatory SED Analyser (VOSA; \citealt{2008A&A...492..277B}). 
Following the construction of the SEDs, best-fit parameters such as effective temperature (T$_{eff}$), luminosity (L), and radius (R) are derived by performing a reduced chi-square ($\chi^2_{red}$) minimisation fitting, which compares the observed fluxes to the synthetic photometric fluxes. The SED fitting technique is described in more detail in \cite{2021ApJ...923..162R}. In addition to the $\chi^2_{red}$ value, VOSA computes two additional parameters, Vgf and Vgf$_{b}$, which correspond to modified $\chi^2_{red}$ values and are used to evaluate the goodness of the fit with minimum fractional residual, especially when the observational flux errors are minimal. For an SED fit to be reliable, the Vgf$_{b}$ value must be less than 15 \citep{2021MNRAS.506.5201R}. To estimate errors in the derived parameters, VOSA makes use of Markov chain Monte Carlo (MCMC) approach.

We used Kurucz stellar atmospheric models \citep{2003IAUS..210P.A20C} to compute synthetic SEDs across a wavelength range extending from FUV to the IR, encompassing observed photometric data points. These models provide the grid of parameters such as T$_{eff}$, metallicity (\big[Fe/H\big]), and surface gravity log \textit{g}.
We fixed metallicity, \big[Fe/H\big] of -0.5 dex, close to the cluster metallicity \citep{2010A&A...517A..32P}. We adopted a range of log\,\textit{g} values from 3$-$5 dex, as Be stars span a range of evolutionary phases \citep{1976IAUS...70...69P} and kept T$_{eff}$ as a free parameter (3,500-50,000 K) in the case of the Kurucz models.
We combined three FUV and two NUV UVIT photometric data points with \textit{Gaia} EDR3 (3 passbands), SLOAN/SDSS (5 passband), 2MASS (3 passbands), and WISE (4 passbands) to construct the observed SEDs. First, we fitted the observed SED, incorporating all the data points, with a single model SED in order to check the overall fit. Out of 23 stars, four stars are well fitted with a single model SED fit, and 19 stars show a significant deviation ( $>$50\%) from the model at shorter wavelengths (FUV region).
To further confirm the UV excess, we repeated the single model SED fit by excluding the data points with wavelengths less than 3000 \AA. After obtaining the stellar parameters, we employed the VOSA binary fit tool to fit the hotter part of the SED for 19 stars. The VOSA binary fitting tool simultaneously fits the hotter and cooler components of a binary system by assuming that the observed flux is a linear combination of two different models. The Kurucz and TMAP atmospheric models \citep{2003ASPC..288..103R} are used to fit the hotter component. The TMAP model grid spans a range of atmospheric parameters, with T$_{eff}$ between 50,000 and 190,000 K, surface gravities (log \textit{g}) between 5 and 9 dex, and a hydrogen mass fraction of zero. For the star GG 101, we were unable to obtain a satisfactory fit using either the Kurucz or TMAP models. Therefore, we adopted the Levenhagen model \citep{2017ApJS..231....1L}, which covers T$_{eff}$ values from 17,000 to 100,000 K and log \textit{g} between 7 and 9.5 dex.

Out of the 19 stars with UV excess, 16 stars are well fitted with a double-component SED, except for three stars: D01 034, GG 109, and SAN 28. We show the best single fit and binary fit SED of Be star GG95 (left panel), and VES 619 (right panel) in Fig.~\ref{sed fit}. The best-fit parameters of the 23 Be stars (seven single and 16 binary) derived from the best SED fits with the lowest $\chi^2_{red}$ and Vgf$_{b}$ values are listed in Table~\ref{comb_table}. We obtained Vgf$_{b}$ values below 15 for all of the Be stars, indicating good SED fits and confirming the reliability of all the derived fundamental parameters. The remaining best-fit single and double-component SEDs are provided in Appendix B. 
Since classical Be stars exhibit excess emission in the near-UV (NUV) and IR due to free-bound and free-free processes occurring in circumstellar discs, respectively \citep{1986MNRAS.222..121G}, these mechanisms are not efficient at producing significant flux in the FUV regime. Also, note that chromospheric activity cannot account for this UV excess, as B-type stars do not exhibit such activity due to the absence of convective envelopes \citep{2008LRSP....5....2H}. Thus, the observed FUV excesses in our sample suggest an additional hot companion. Furthermore, to assess the evolutionary stage and nature of hot companions, we place them in the Hertzsprung-Russell diagram (HRD), as discussed in the subsequent section.

\begin{figure}
    \centering
    \includegraphics[scale=0.38]{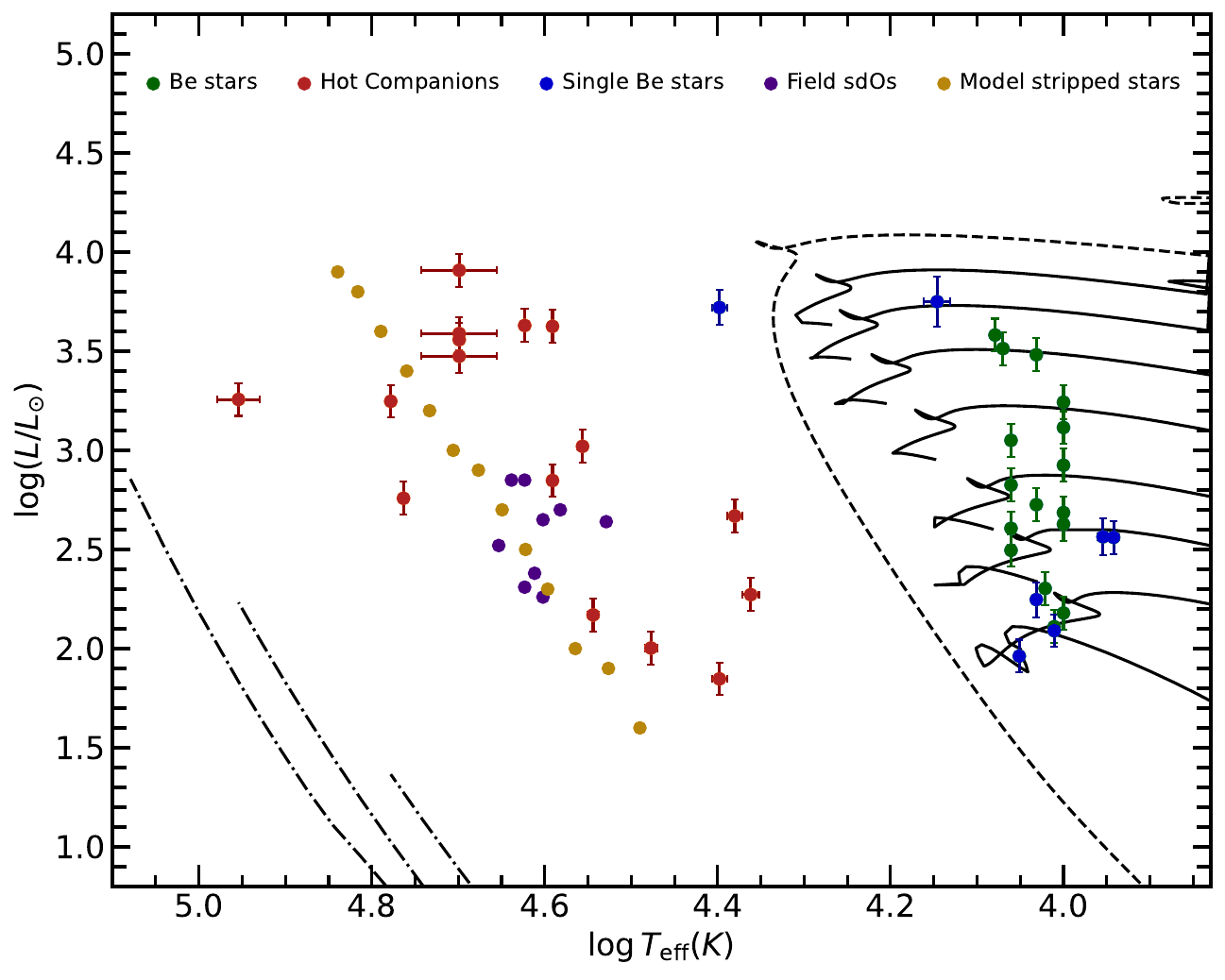}
    \caption{HR diagram illustrating UVIT-identified Be stars and their hot companions. The cluster isochrone (dashed black line), WD evolutionary tracks (dash-dotted black line, 0.3–0.5 M$\odot$, from right to left), and MIST tracks (solid black line, 3.0–8.2 M$_{\odot}$, bottom to top) are shown for comparison. Cool and hot components of binary Be stars are marked in green and brown, and single-fit Be stars are in blue. Field Be+sdO systems from \citet{2021AJ....161..248W} are displayed in purple, and model stripped stars with masses of 0.58, 0.66, 0.74, 0.85, 0.97, 1.11, 1.27, 1.43, 1.63, 1.88, 2.17, 2.49, and 2.87 M$\odot$ from \citet{2018A&A...615A..78G} are shown as golden circles (bottom to top).}
    
    \label{HR_dia_other_stripped_stars}
\end{figure}

\section{Discussion}
\label{discussion}

This is the first study to trace possible binary companions of Be stars in young clusters. NGC~663 is known to harbour a large number of Be stars for a very long time. These stars also have a large range in spectral type, leading to the inference that the 'Be-phenomenon' is operational across the B spectral type in this cluster. Modelling the UV excess and the multi-wavelength SED of 23 Be stars has led to the detection and characterisation of binary companions in 16. The hot companions, as shown in HRD, (Fig.~\ref{HR_dia_other_stripped_stars}) belong to the sdOB spectral types and appear to be similar to those found in the field. This study, therefore, provides the first evidence of a binary pathway to Be formation in a young Galactic cluster. NGC~663 has at least 69.5\% of Be stars formed through binary mass-transfer, suggesting that a majority of Be stars in NGC~663 are formed via binary mass transfer.

\cite{2021AJ....161..248W} detected spectral signatures of the sdO companions in ten out of 13 Be stars, and a comparison of their temperatures and radii with evolutionary tracks indicates that the sdO stars occupy the relatively long-lived, He-core burning stage. The companions identified to Be-stars in NGC~663 also occupy a very similar space in the HR diagram (see Fig. 17 of \citealt{2021AJ....161..248W}), suggesting that they are sdO stars in the long-lived He-core burning phase.

The cluster NGC~663 is 25 Myr old and therefore has a turn-off mass of $\sim$ 9.2 M$_\odot$. Recently, \citep{2024AJ....167...12C} estimated the cluster's age to be 39 Myr, which corresponds to a turn-off mass between $\sim$ 7$-$8 M$_\odot$. All 16 sdOBs likely formed within 25 Myr from a primary star mass of 7-9 M$_\odot$ in a binary system. Even though the initial mass is similar, the end mass appears to be different. When we compare the L and T$_{eff}$ of the detected sdOBs with the models from \cite{2018A&A...615A..78G}, we find that the mass of the end products likely ranges from 0.6 - 2.8 M$_\odot$. The primary stars would have lost different amounts of mass during mass transfer, possibly due to different orbital parameters. The Be stars, which are the cooler companion of the sdOBs, also have a mass range of approximately 2-8 M$_\odot$, suggesting that the Be binaries have a wide range of primary-to-secondary star mass-ratios. This likely explains the appearance of Be phenomenon across the B spectral type in this cluster. The dynamical environment of the cluster also likely plays a role in binary evolution, as this can alter binary orbits. The orbital information for these systems is currently unknown, except that many are known to be photometric and/or spectroscopic variables. The binary Be systems in this cluster are therefore very important for investigating binary mass transfer mechanisms in massive stars. The sdOB companions detected here put strong constraints on the mass transfer mechanisms that are crucial to theoretical models such as those presented in \cite{2018A&A...615A..78G}. 

The high-resolution spectroscopic studies of the identified Be binary systems in NGC~663 need to be performed to confirm the evolutionary status as well as to explore the atmospheric abundances of the sdOBs. Photometric and spectroscopic variability of these systems also need to be studied systematically to identify orbital parameters. In the future, we plan to extend this study to more young clusters harbouring Be stars.

\section{Summary}
\label{summary}

The main results from this work can be summarised as follows:
\begin{itemize}
    \item We utilised UVIT/AstroSat observations in 3 FUV and 2 NUV filters to characterise 23 Be stars in the young OC NGC~663. 

    \item We estimated atmospheric parameters such as T$_{eff}$, Ls, and Rs of the Be stars by fitting multi-wavelength SEDs with a single model spectrum. Among the 23 Be stars considered, 19 exhibited significant UV excess.
    
    \item Sixteen Be stars were well fitted with a double-component SED, indicating the presence of a hot companion. The nature of these hot companions was assessed by placing them on the HR diagram and comparing their position with the theoretical evolutionary tracks or previously observed field stars of similar class. The positions of the detected hot companions coincide with the location of stripped stars  \citep{2021AJ....161..248W}, suggesting that the companions are likely stripped stars belonging to the sdOB type in the long-lived He-core burning phase.

    \item The estimated masses of the hot companions range from 0.6 to 2.8 M$\odot$, based on the models of \citet{2018A&A...615A..78G}, while the Be stars have masses between 2.0 and 8.0 M$\odot$, derived from comparison with MIST evolutionary tracks, with many of them being slightly evolved from the MS.
    
    \item This study establishes the binary pathway for the formation of Be stars in young clusters, with 69.5\% of Be stars in NGC~663 found in binary systems. Be binary systems are interesting targets for further high-resolution spectroscopic studies to probe their evolutionary status and to gain insights into the abundances of the sdOB stars. 

    \end{itemize}

\begin{acknowledgements}

We thank the referee for providing the productive report that improved the quality of our manuscript. AS acknowledges support from the SERB Power Fellowship. SN and SR acknowledges support from StarDance: the non-canonical evolution of stars in clusters (co-funded by the European Union, ERC-2022-AdG, Grant Agreement 101093572, PI: E. Pancino). SN also extends gratitude to Blesson Mathew, Sipra Hota, Ashish Devraj, Akhil Krishna and Vikrant Jadhav for the fruitful discussions.
This publication utilises data from the \textit{AstroSat} mission’s UVIT, archived at the Indian Space Science Data Centre (ISSDC). The UVIT project is a collaboration between IIA Bengaluru, IUCAA Pune, TIFR Mumbai, several ISRO centres, and CSA. This research also made use of VOSA, developed under the Spanish Virtual Observatory project, supported by the Spanish MINECO through grant AyA2017-84089.

\end{acknowledgements}

\textit{Software}: Topcat (\citealt{2011ascl.soft01010T}), Matplotlib (\citealt{4160265}), NumPy (\citealt{5725236}), Scipy (\citealt{4160250}, \citealt{5725235}), Astropy (\citealt{refId0},\citealt{2018AJ....156..123A}), and Pandas (\citealt{mckinney-proc-scipy-2010}).


\bibliographystyle{aa}
\bibliography{example}

%%%%%%%%%%%%%%%%% APPENDICES %%%%%%%%%%%%%%%%%%%%%

\onecolumn
\begin{appendix}
\section{Colour-magnitude diagrams}
\label{cmds}

\cite{2023yCat..36730114H} present the Proper motion (PM) membership study of 7167 star clusters using \textit{Gaia} DR3 catalogue \citep{GaiaDR3_VizieR_2022}, including NGC~663, and they found 1081 stars as PM members of the cluster with membership probability (P$_{PM}$) more than 50\%. To identify optical counterparts of UVIT detections, we cross-matched the \cite{2023yCat..36730114H} catalogue (P$_{PM}$>50\%) with sources detected in both FUV and NUV images. The P$_{PM}$ of the 23 Be stars are tabulated in Table \ref{comb_table}.

Figure~\ref{fig:uvcmds} shows the optical, FUV-optical, and NUV-optical colour-magnitude diagrams (CMDs) along with overlaid MIST isochrones taken from \cite{2016ApJ...823..102C} generated for both UVIT and \textit{Gaia} DR3 filters for the cluster parameters (as mentioned in the caption).
We used isochrones with a metallicity of [Fe/H] = -0.125 dex and Z = 0.003110, excluding initial rotation effects. The optical magnitudes were converted from Vega to AB magnitude system using appropriate conversion factors provided in \textit{Gaia} documentation\footnote{https://www.cosmos.esa.int/web/gaia-users/archive}. 

We cross-matched UVIT-detected stars with Be stars catalogued in \cite{2015AJ....149...43Y}, identifying 23 stars. Of these, 22 are detected in F148W, F154W, and F169M, 1 in N263M, and 23 in N279N, all with membership probabilities exceeding 50\% . All identified Be stars are represented by green triangles, cyan rectangles and blue pentagons in the optical CMD, as illustrated in Fig.~\ref{fig:uvcmds}. A significant spread is observed along the MS, especially at its bright end spanning $\sim$0.45\,mag in BP$-$RP colour. The FUV-optical and NUV-optical CMDs also show the spread at the bright end of the MS, but FUV and NUV optical colours span $\sim$2\,mag larger than optical colour mainly due to the sensitivity of the UV-optical colour to T$_{eff}$. Also, the turn-off region of NGC~663 appears slightly broadened, partially influenced by the presence of Be stars. Extended main-sequence turn-offs (eMSTOs), likely caused by enhanced stellar rotation, have been observed in OCs with ages between approximately 50 Myr and 2 Gyr (see \citep{2024MNRAS.532.1547C} and references therein). However, NGC~663 is younger than the typical eMSTO clusters, and the spread at the turn-off appears to be prominent; therefore, this  cluster may be the youngest to show the eMSTO phenomenon. As this cluster is known to have a differential reddening, we also adopt an extinction range from  A$_{v}$=2.17 to A$_{v}$=2.9 in all CMDs. 

Since the Be stars have a high rotational velocity, their T$_{eff}$ are affected leading to fast rotators on the red edge of the MS, whereas slow rotators in its blue side. To assess the impact of T$_{eff}$ on UV-optical colour (which is more sensitive to T$_{eff}$), colour-colour diagrams (CCDs) are constructed using the FUV and NUV filters. The linear plot of (Gbp-Grp) against (F154W-G), (F169M-G), and (N279-G) are shown in Fig.~\ref{fig:uvcmds}. In all CCDs, optical colour (Gbp-Grp) spans a 0.6\,mag range, whereas the FUV-G colour spans about 3 mag, a similar trend can be seen in the case of the NUV-G ($\sim$2.5\,mag) colour diagram. We note that some Be stars show bluer UV-optical colour, suggestive of a possible UV excess, as they are bluer in the UV-optical colour.

\begin{table}[htbp]
\centering
\caption{UVIT observations and different extinction values adopted for NGC~663.}
\label{uvit_combined}
\begin{tabular}{llllllllllll}
\hline
Filter & $\lambda_{\rm mean}$ (\r{A}) & $\Delta\lambda$ (\r{A}) & ZP (AB mag) & T$_{\rm exp}$ (sec) & A$_{V}$ & A$_{\rm F148W}$ & A$_{\rm F154W}$ & A$_{\rm F169M}$ &  A$_{\rm N263M}$ & A$_{\rm N279N}$ \\
 \hline
 \hline
 \noalign{\vskip 2mm}
F148W  & 1481 & 500 & 18.09 & 1191.673 &  2.17$^{a}$   & 5.794 & 5.664 & 5.555 & 4.579 & 4.231 \\
F154W  & 1541 & 380 & 17.77 & 493.823 &  2.268$^{b}$  & 6.056 & 5.919 & 5.806 & 4.785 & 4.423 \\
F169M  & 1608 & 290 & 17.41 & 194.791 &  2.79$^{c}$   & 7.449 & 7.282 & 7.142 & 5.887 & 5.440 \\
N263M  & 2632 & 275 & 18.14 & 1201.311 & & & & & &\\
N279N  & 2792 & 90  & 16.50 & 1412.314 & & & & & & \\

\hline
\end{tabular}
\tablefoot{
The first five columns list filter characteristics and total exposure times. The next six columns present extinction values, A$_V$, reported in the literature and the corresponding extinction in UVIT passbands.}
\tablebib{
(a)~\citet{2018AJ....155...50G}; (b) \citet{2021MNRAS.504..356D}; (c) \citet{Angelo_2022}.
}
\end{table}

\begin{figure}[htbp]
    \centering
    \begin{subfigure}[t]{0.46\textwidth}
    \centering
    \includegraphics[height=4.3cm]{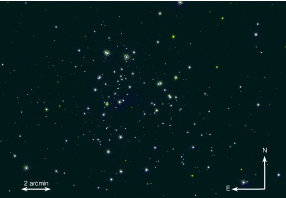}
    \caption{Composite two-colour image of NGC~663: blue (F148W, FUV) and yellow (N279N, NUV).}
    \label{uv_color_image}
    \end{subfigure}
    \hfill
    \begin{subfigure}[t]{0.48\textwidth}
        \centering
        \includegraphics[height=4.3cm]{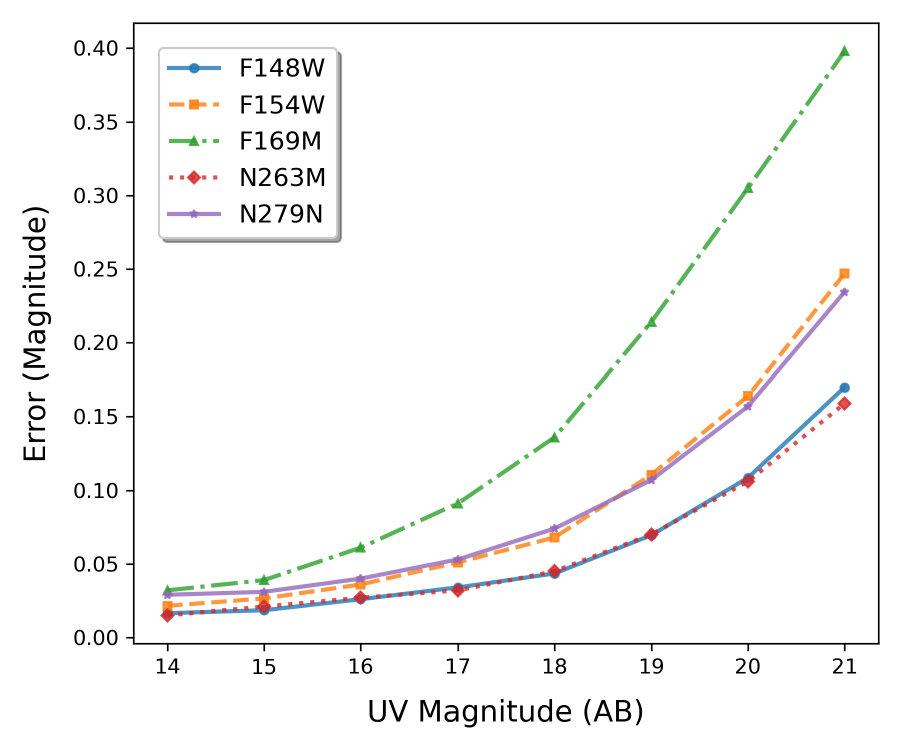}
        \caption{PSF-fit errors (mean) vs magnitude for UVIT observations of NGC~663 in three FUV and two NUV bands.}
        \label{psf_fit_err}
    \end{subfigure}
    \caption{Left: UVIT two-colour image of NGC~663. Right:  PSF-fit magnitude errors.}
    \label{fig:side_by_side}
\end{figure}

\begin{figure}
    \centering
    \includegraphics[scale=0.35]{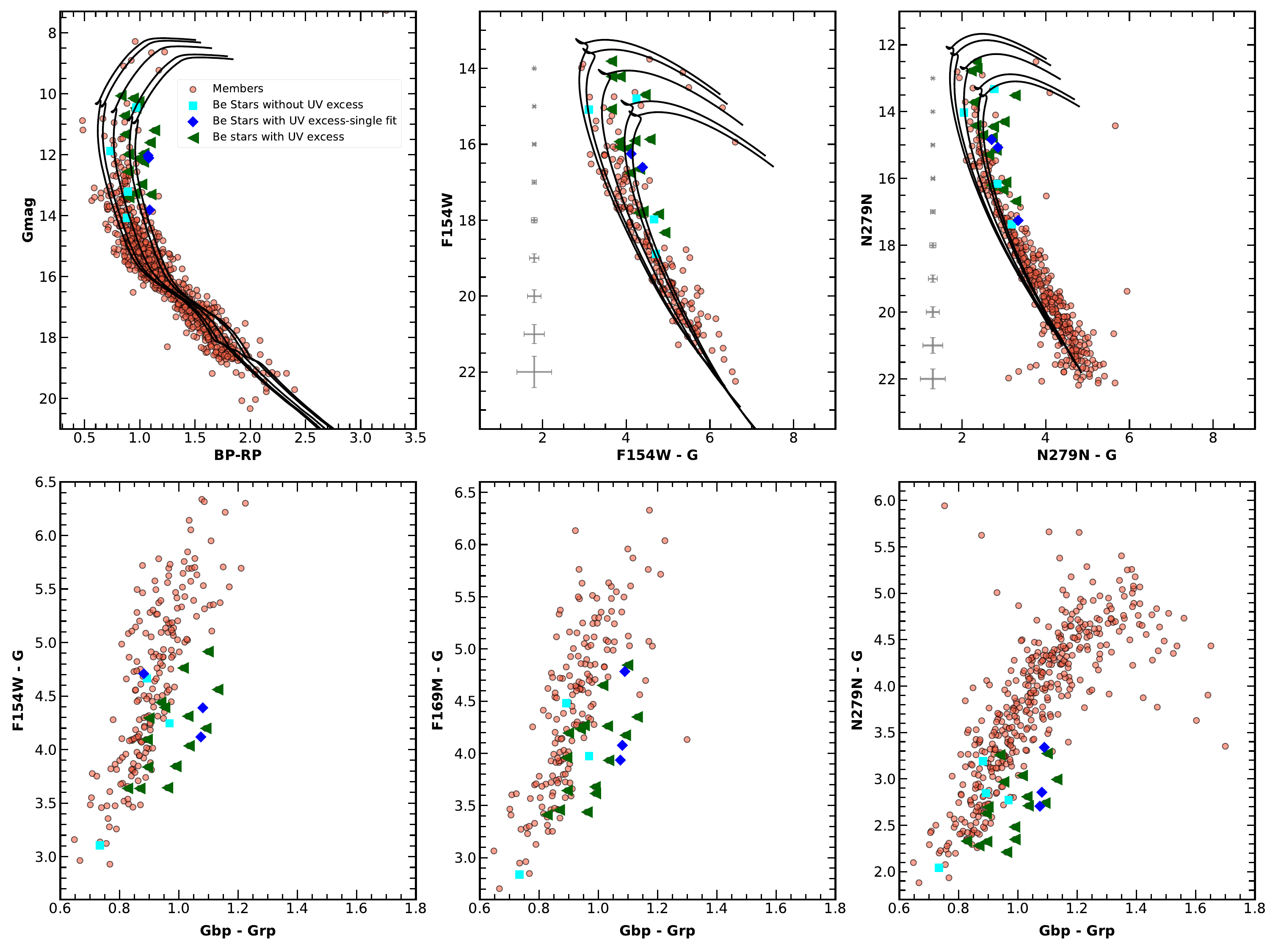}
    \caption{Top panel: Optical CMD of NGC~663, based on \textit{Gaia} DR3 photometry (first panel).
    The red points represent the optical counterparts (\textit{Gaia} DR3) of the UVIT FUV/NUV detections. Be stars with UV excess, Be stars without UV excess, and Be stars showing UV excess but not fitted with a double component SED are shown as green triangles, cyan squares, and blue pentagons, respectively. The MIST isochrones corresponding to an age of 25 Myr and metallicity [Fe/H] = −0.125 dex are overplotted as solid black lines for various values of A$_{V}$, ranging from 2.17 to 2.9 (from left to right), assuming a reddening of E(B−V) = 0.7\,mag and a distance modulus of (m−M)$_V$ = 11.86\,mag. The second and third panels in the upper panel show the FUV–optical and NUV–optical CMDs created using the F154W and N279N passbands, respectively, for the confirmed members of the cluster. Median error bars are displayed in grey on the left side of FUV and NUV-optical CMDs.
    Bottom panel: CCDs in the FUV–F154W vs F154W, FUV–F169M vs F169M, and NUV–N279N vs N279N planes, respectively. The meaning of the symbols shown is provided in the top-left panel of the first column.}
    \label{fig:uvcmds}
\end{figure}

\section{SED fits of Be stars}
The details of the SED fitting are elaborated in Sect.~\ref{sed}. The SED fits, single as well as binary, of the rest of the Be stars are shown in Fig. \ref{SEDs_appendix}

\begin{figure*}
    \centering
    \includegraphics[width=0.41\columnwidth]{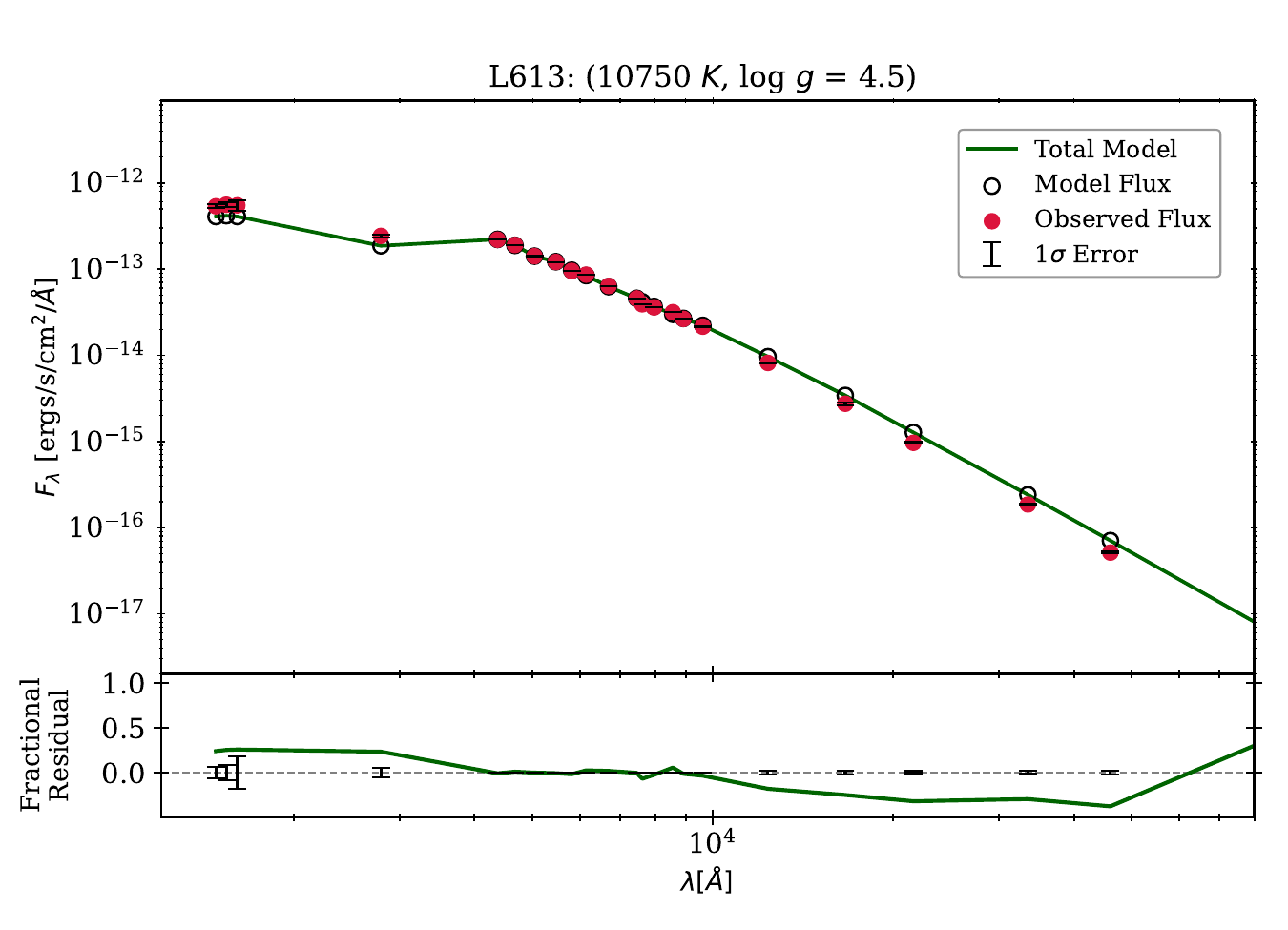}
    \includegraphics[width=0.41\columnwidth]{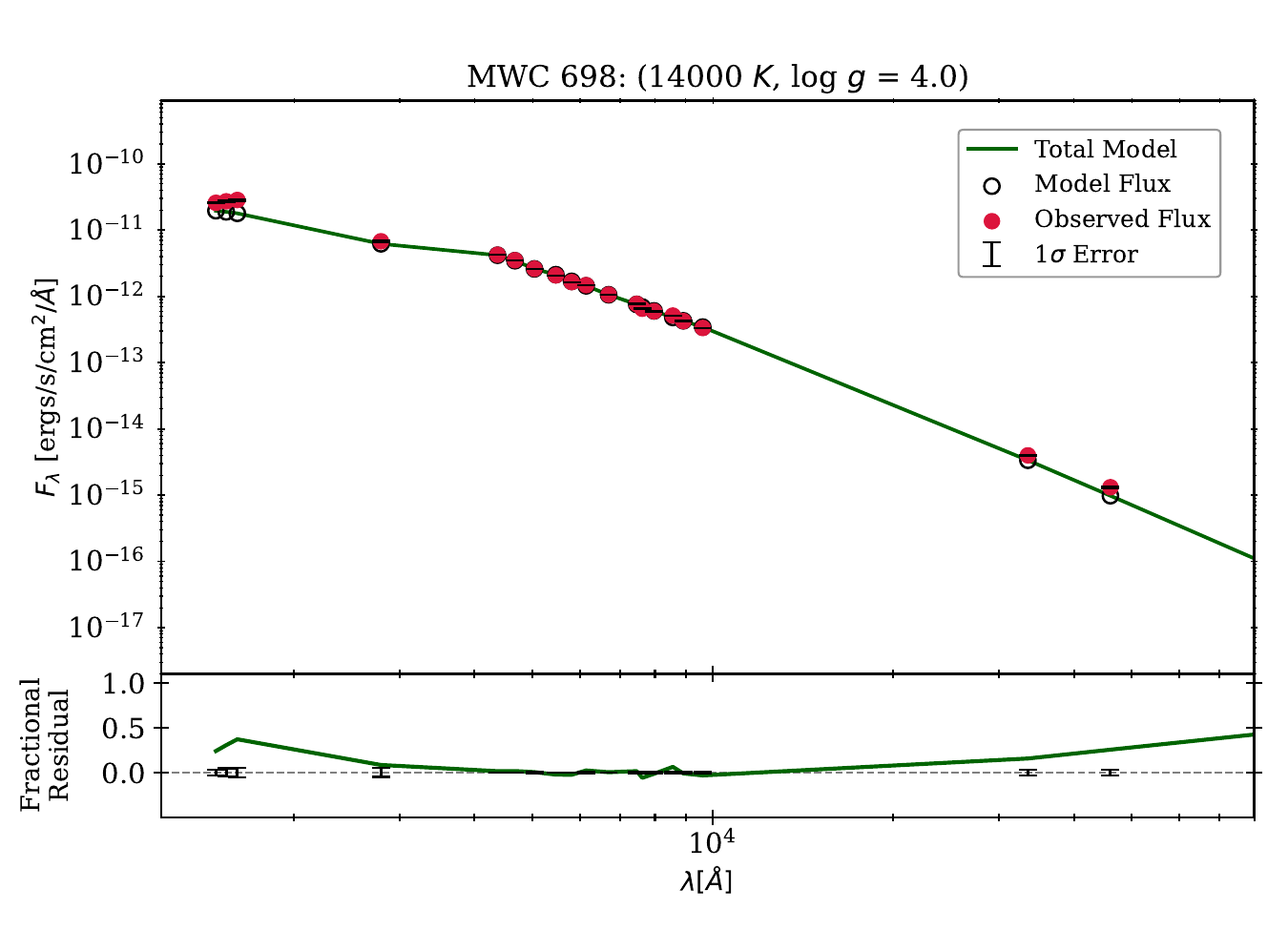}
    \includegraphics[width=0.41\columnwidth]{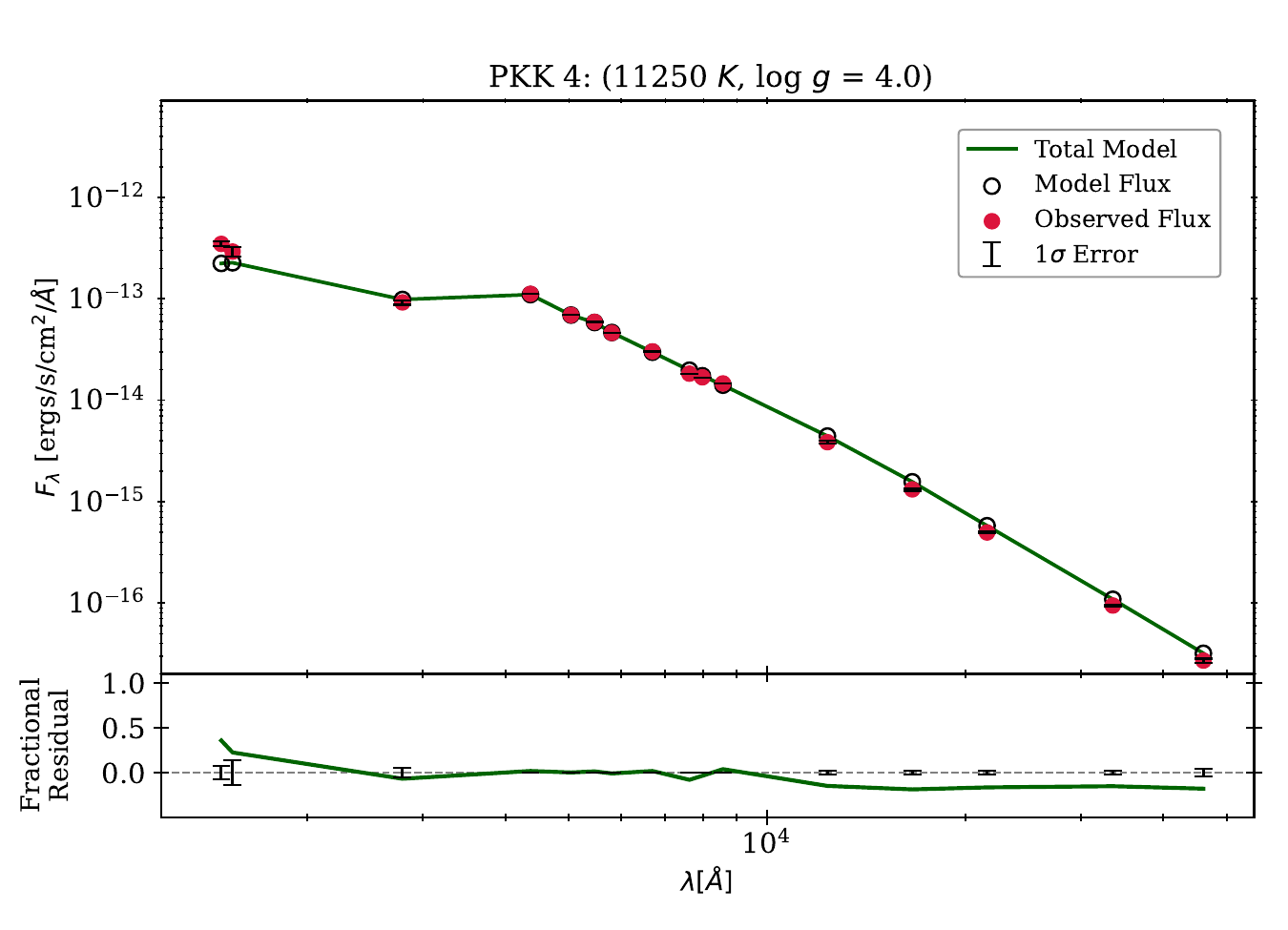}
    \includegraphics[width=0.41\columnwidth]{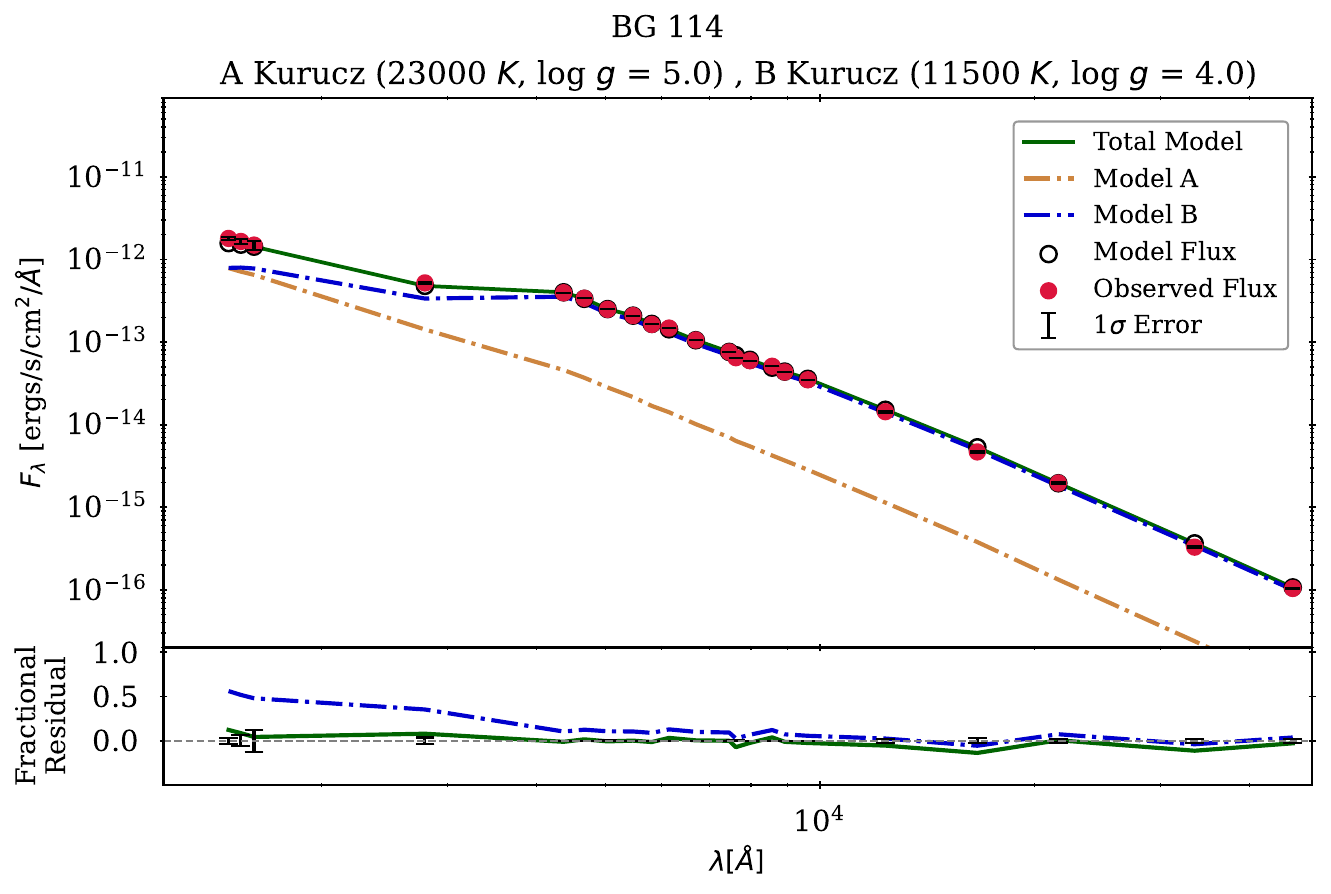}
    \includegraphics[width=0.41\columnwidth]{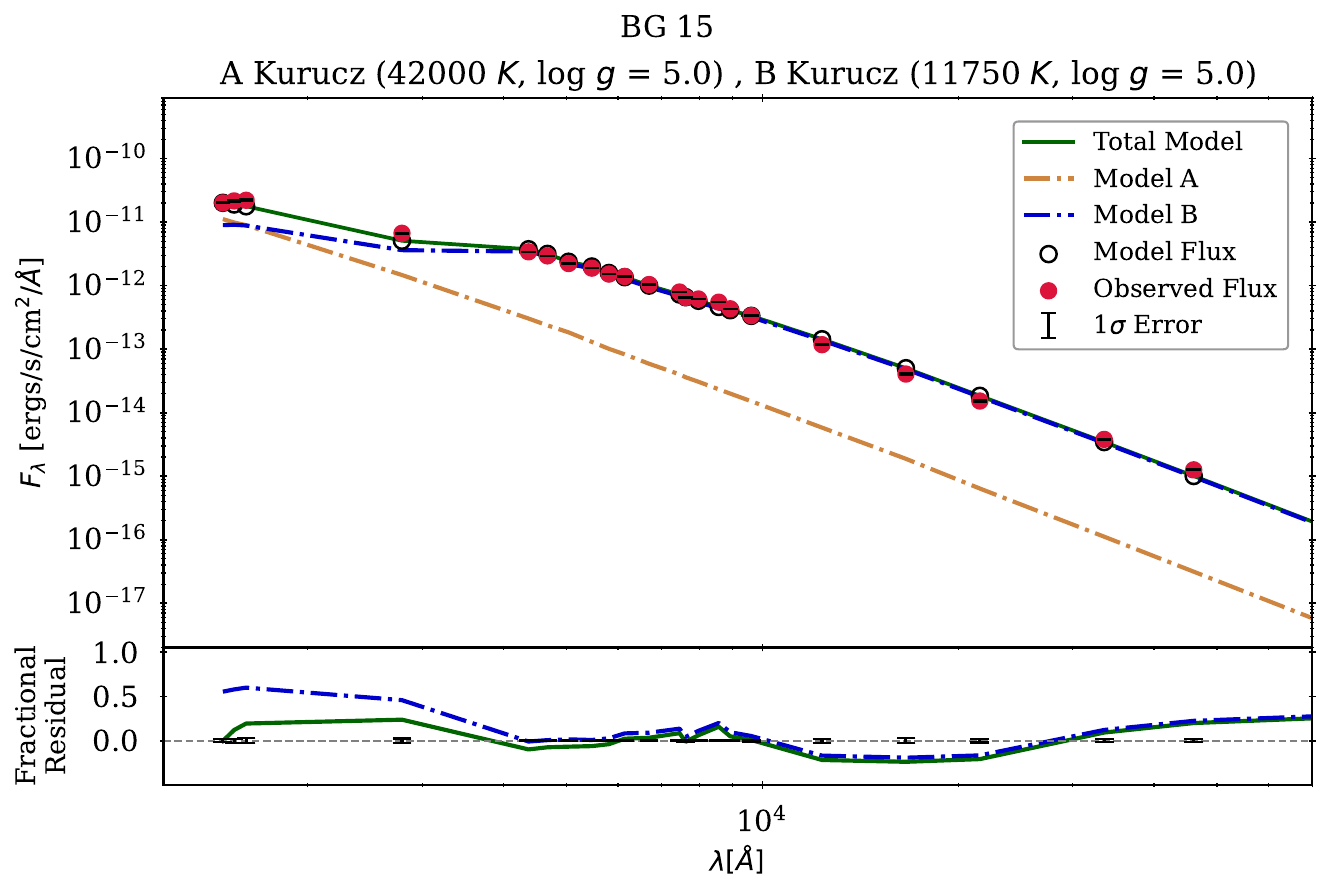}
    \includegraphics[width=0.41\columnwidth]{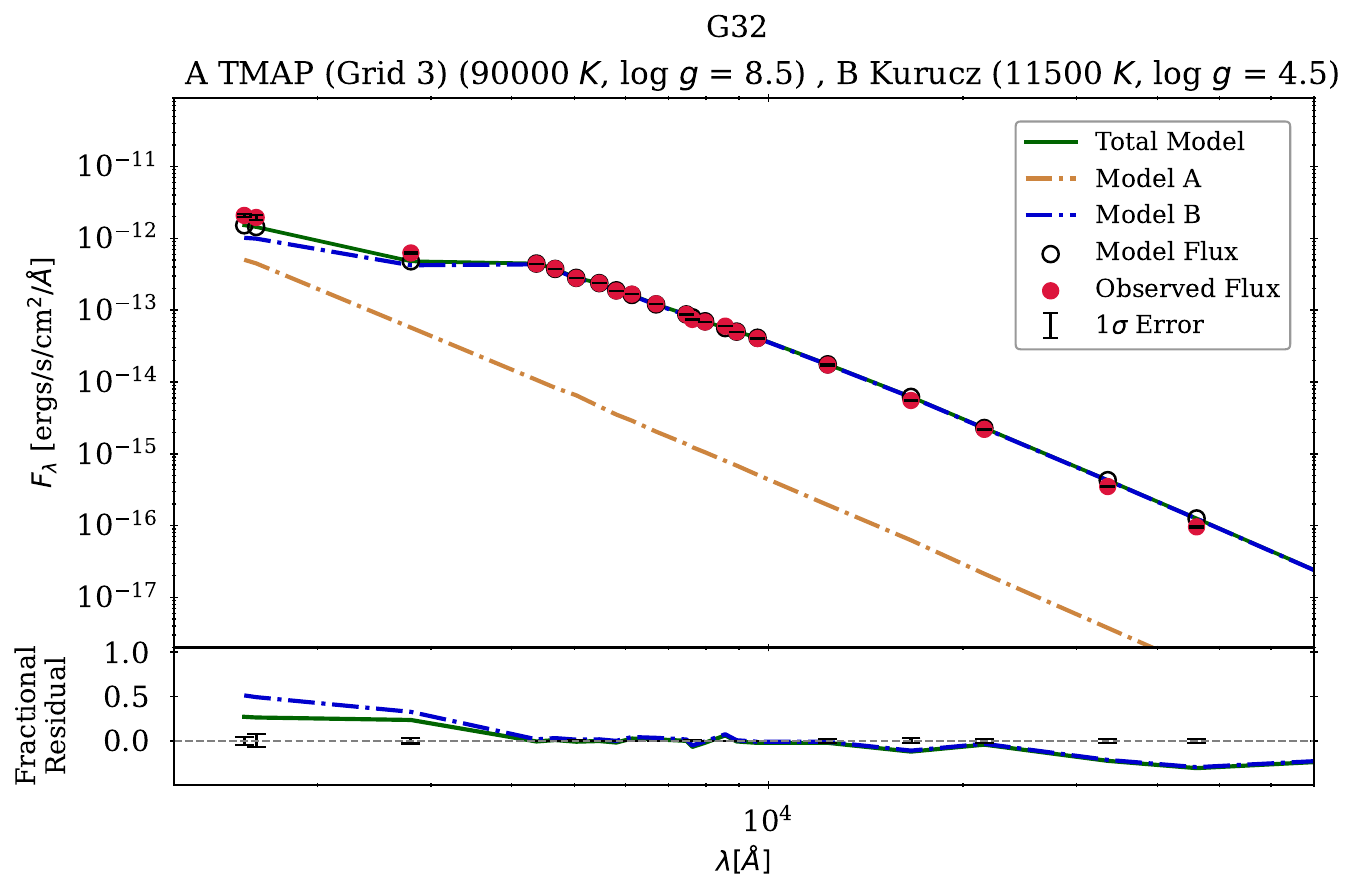}
    \includegraphics[width=0.41\columnwidth]{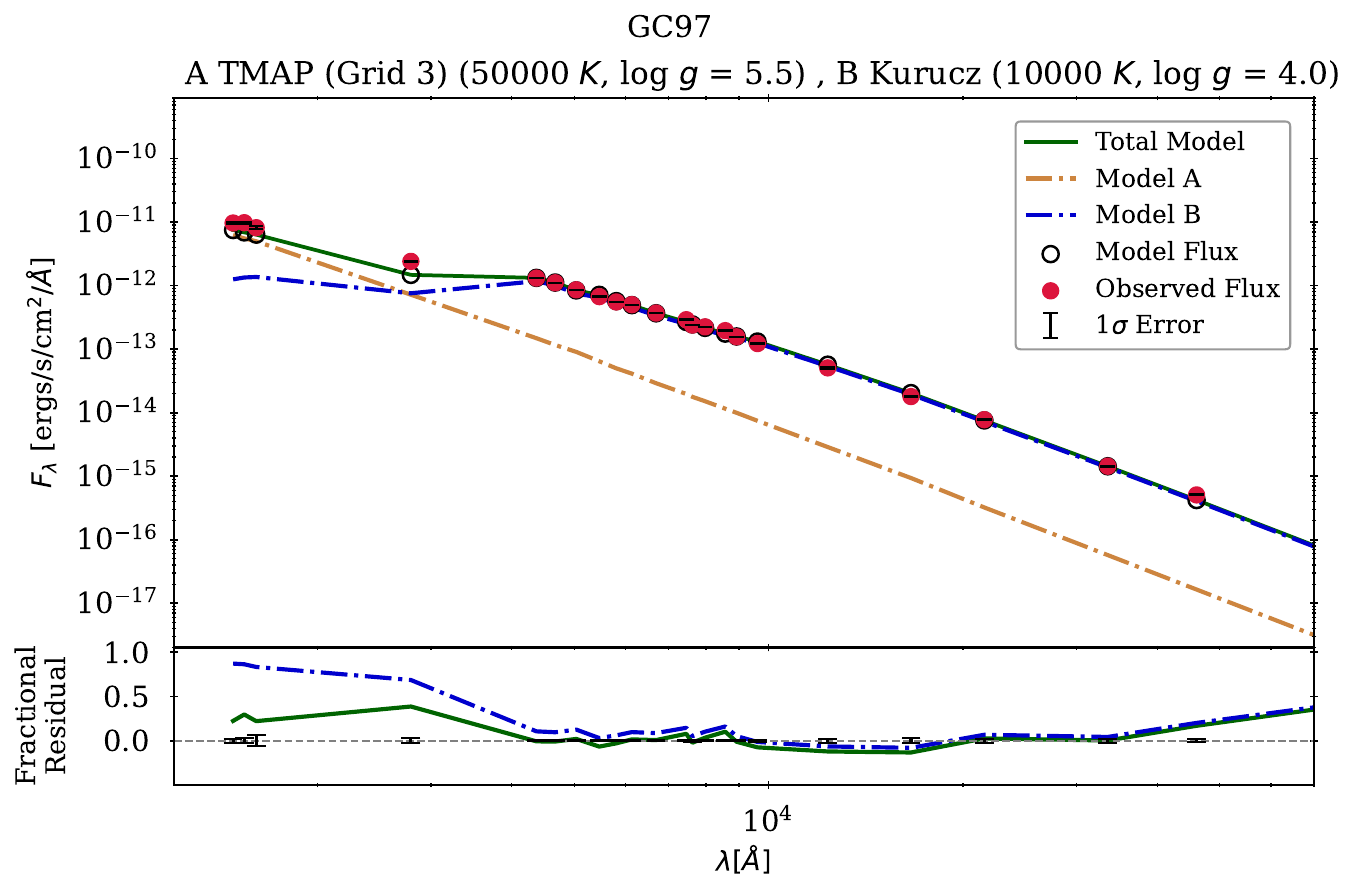}
    \includegraphics[width=0.41\columnwidth]{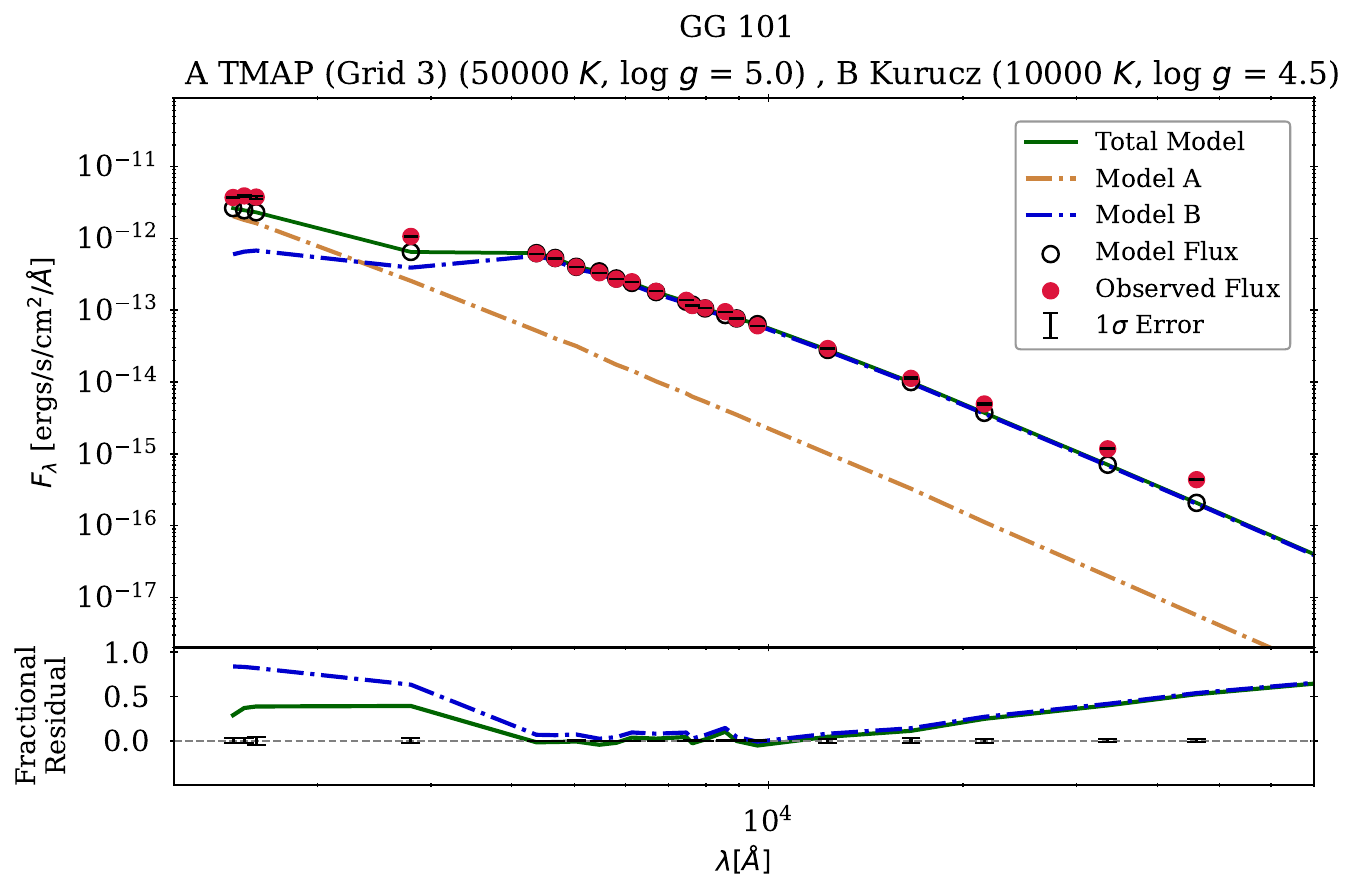}
    \caption{Best SED fits, including single and binary model fits, of the remaining Be stars. The SEDs illustrated in the uppermost panel consist of three stars showing no UV excess. They are well fitted with a single model SED. The symbols and models are the same as in Fig.~\ref{sed fit}.}
    \label{SEDs_appendix}
\end{figure*}

\begin{figure*}
    \ContinuedFloat
    \centering
    \includegraphics[width=0.41\columnwidth]{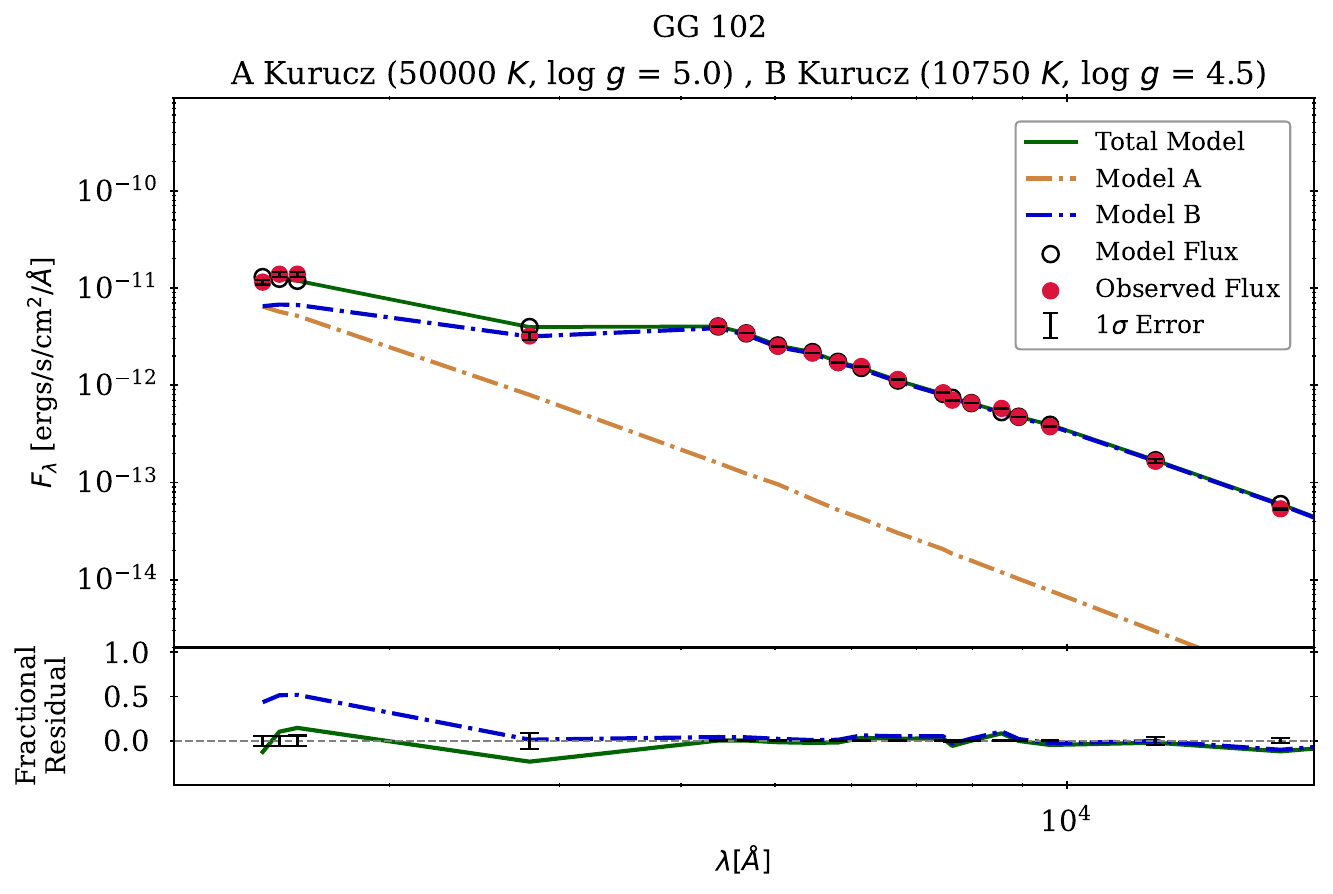}
    \includegraphics[width=0.41\columnwidth]{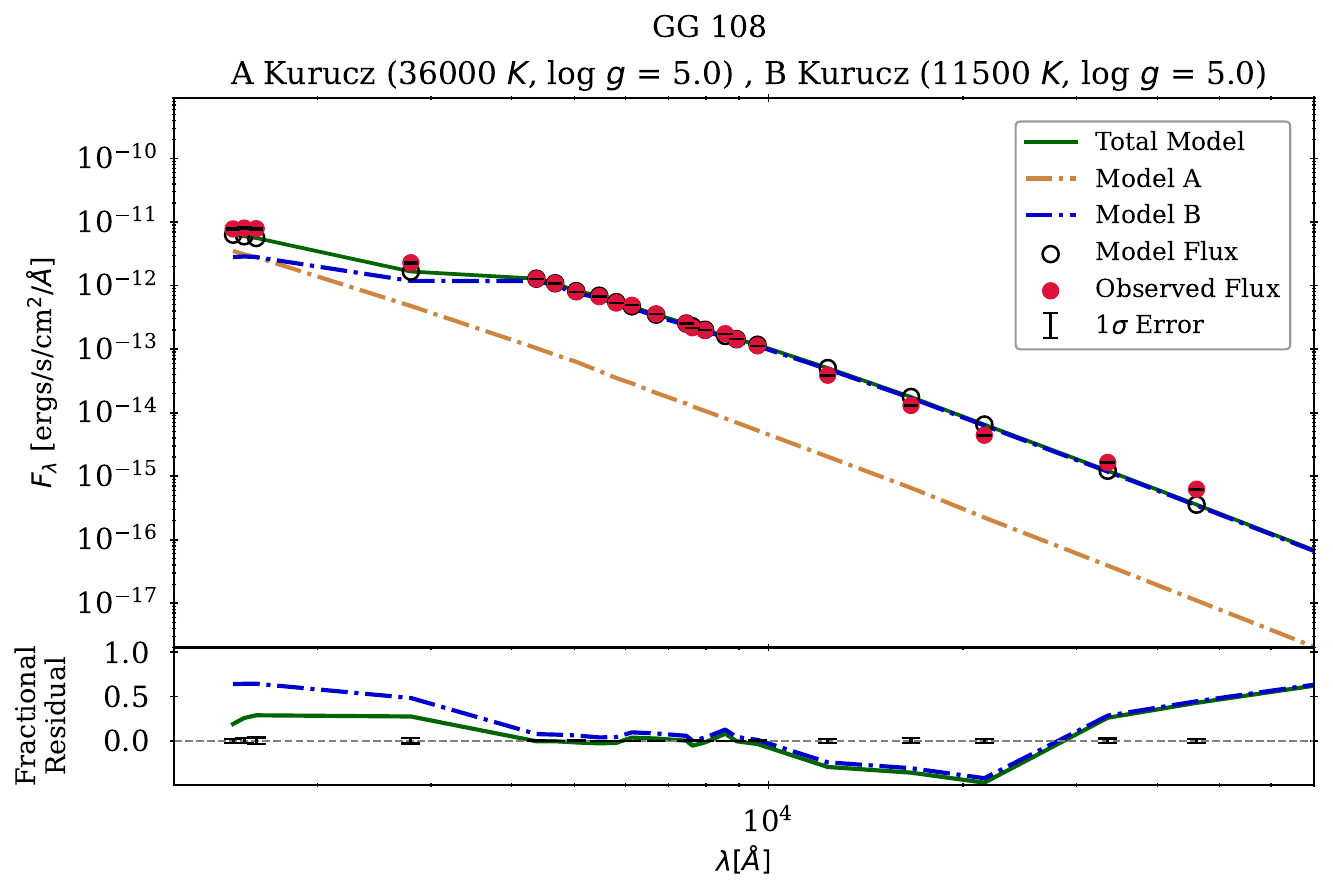}
    \includegraphics[width=0.41\columnwidth]{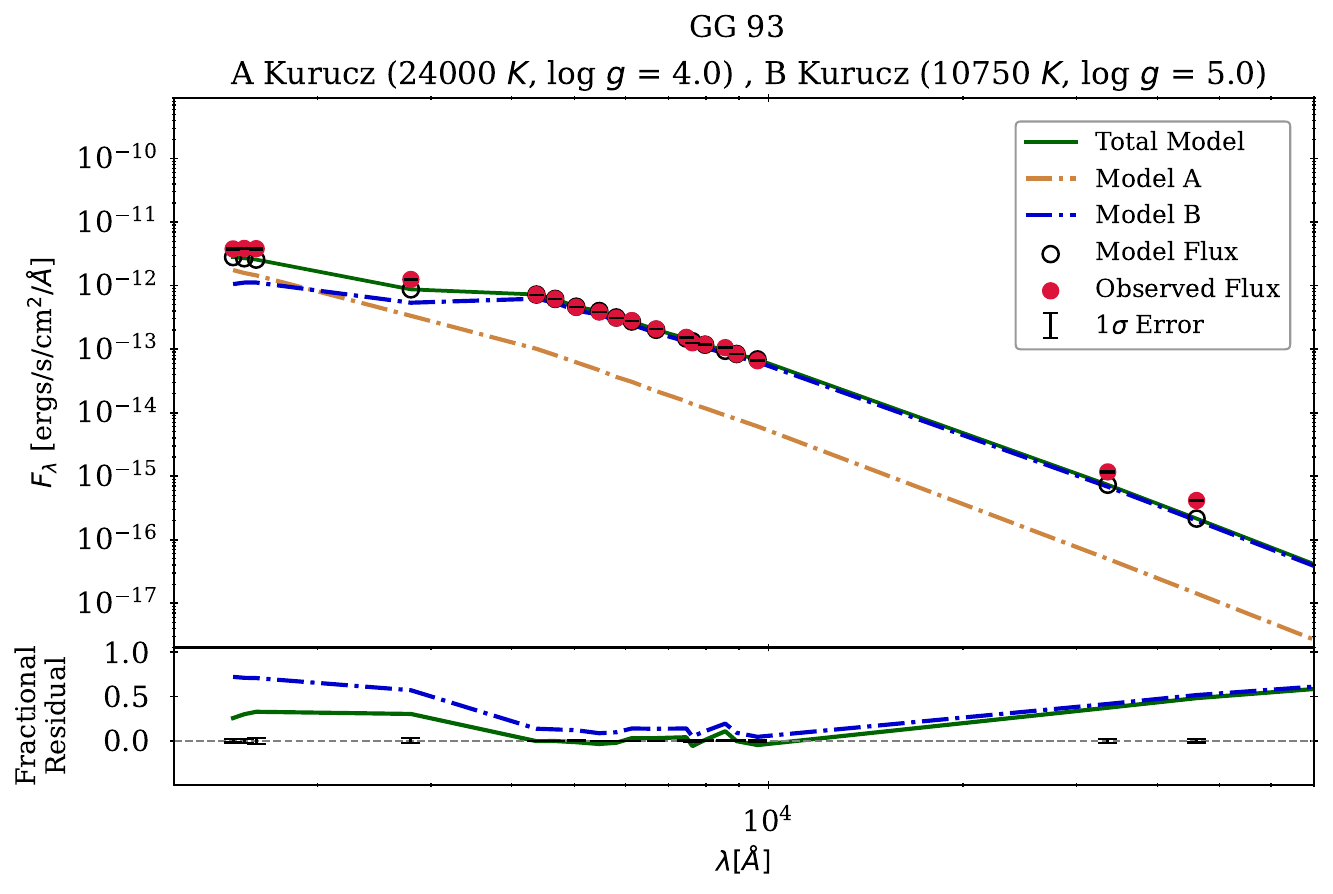}
    \includegraphics[width=0.41\columnwidth]{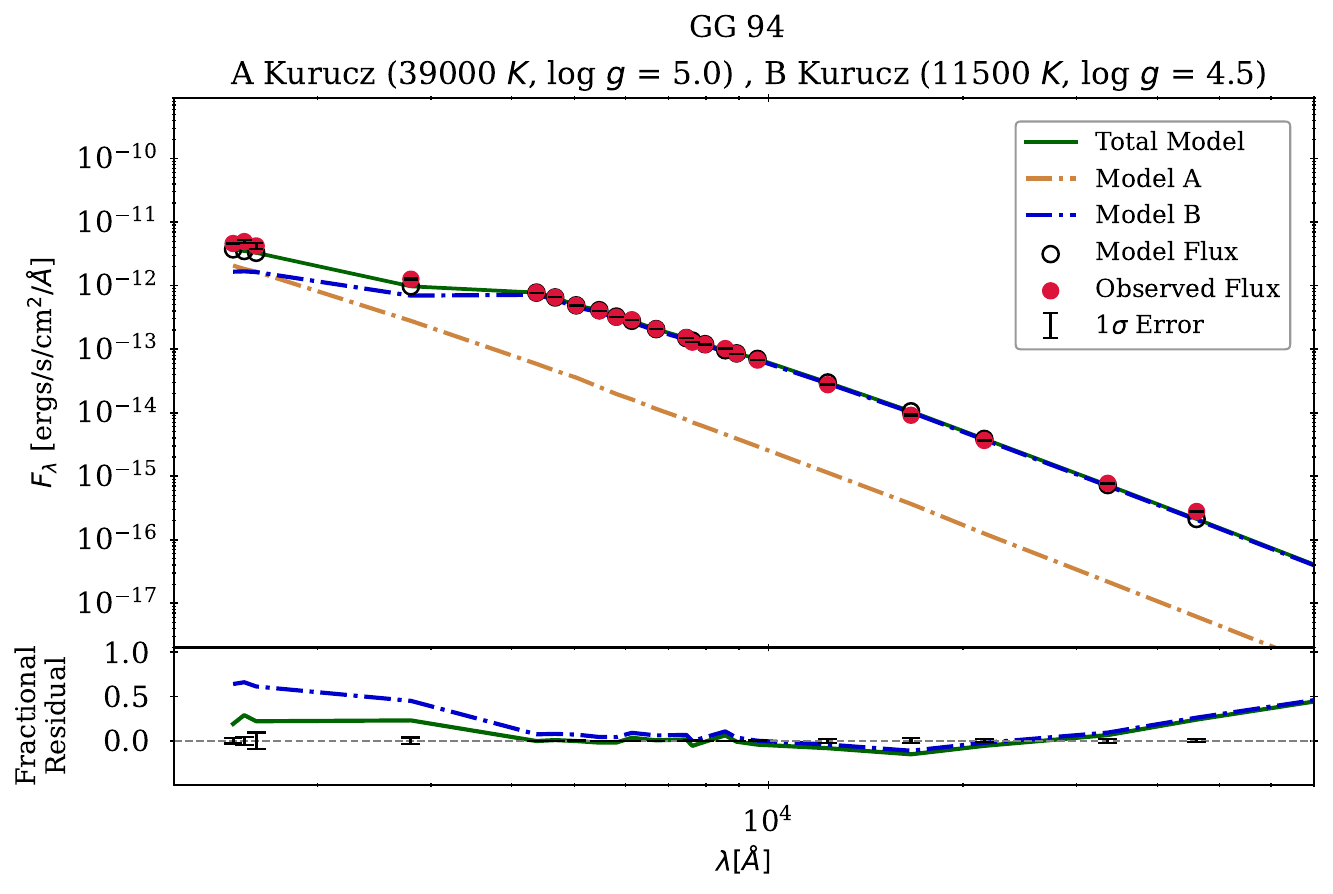}
    \includegraphics[width=0.41\columnwidth]{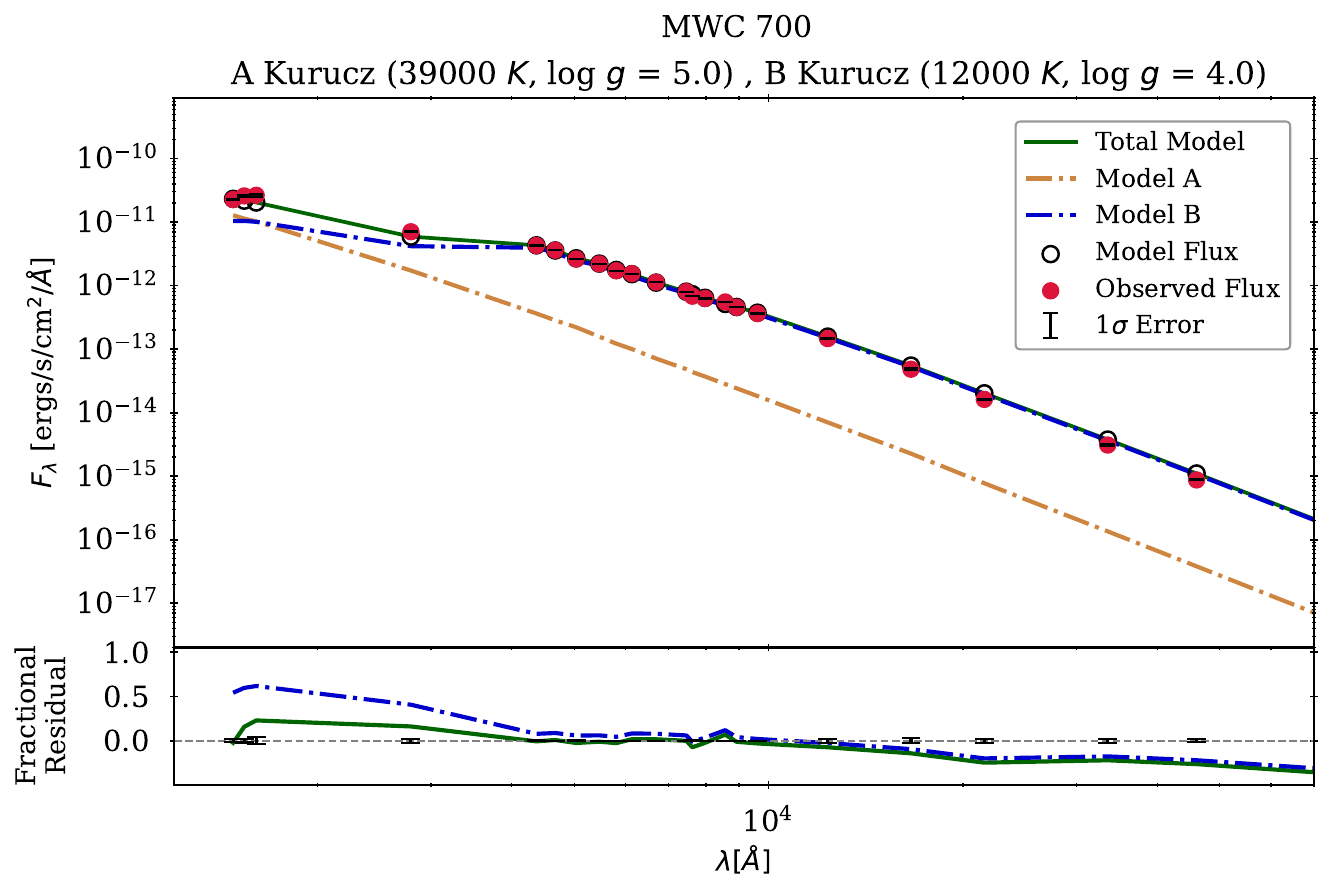}
    \includegraphics[width=0.41\columnwidth]{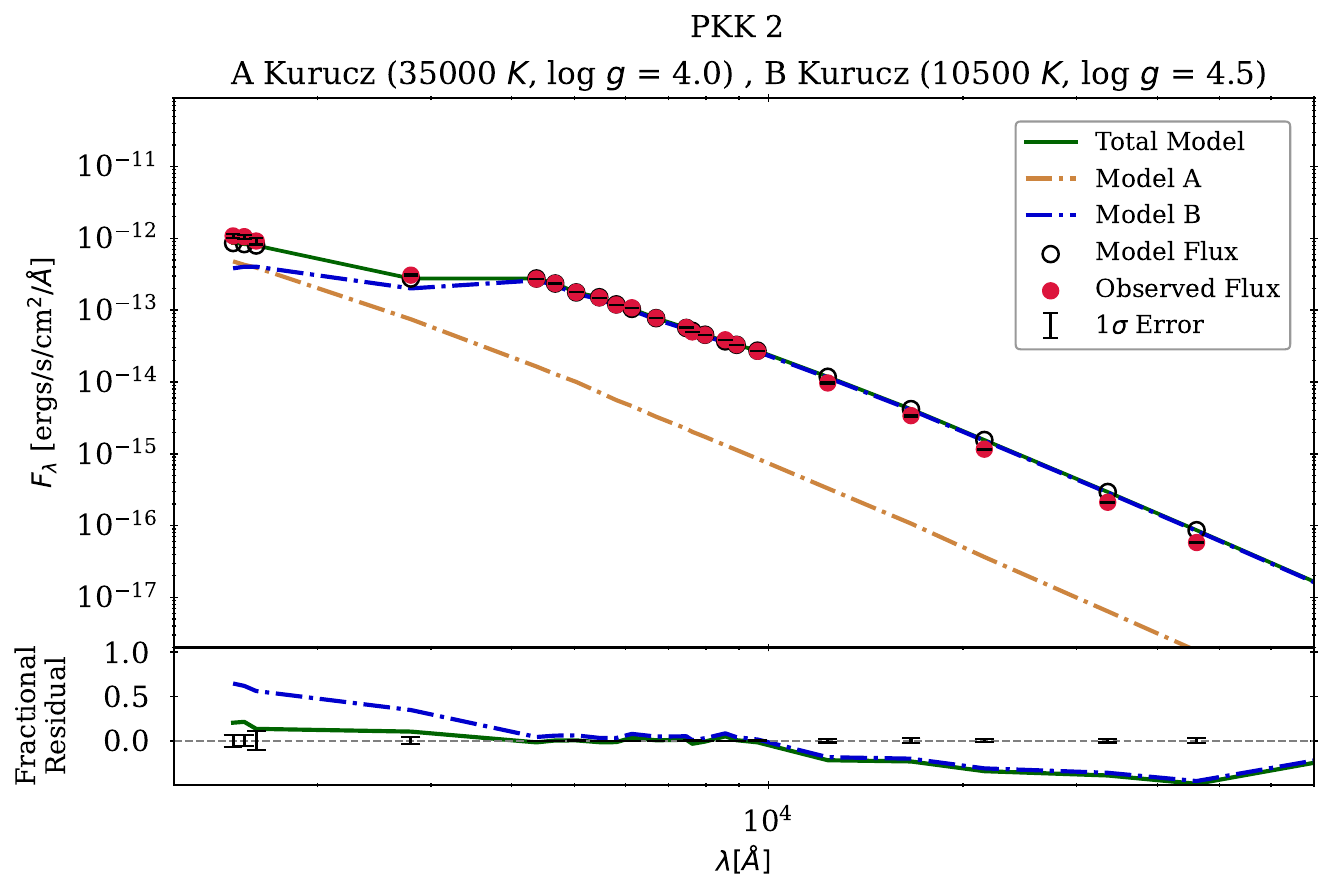}
    \includegraphics[width=0.41\columnwidth]{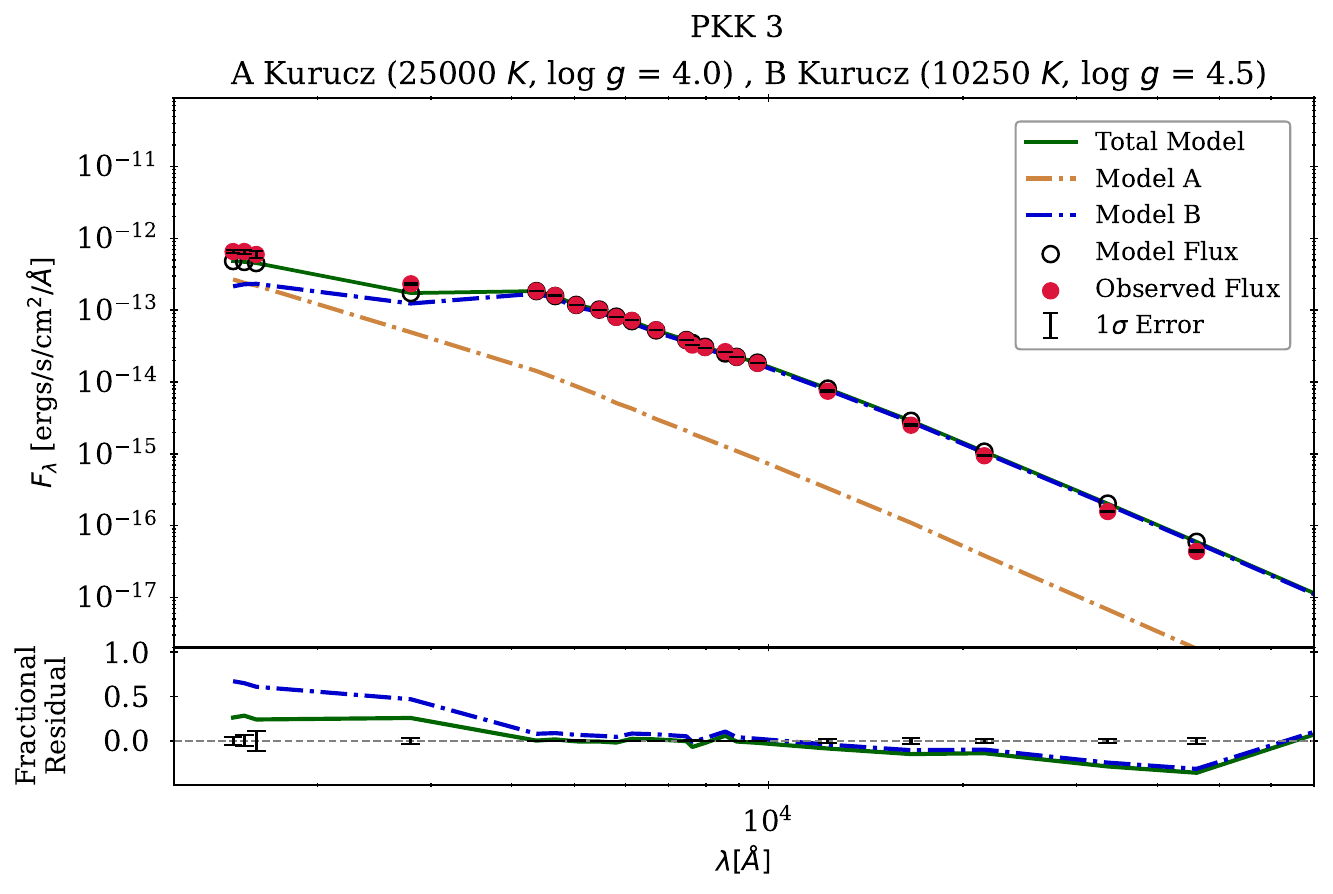}
    \includegraphics[width=0.41\columnwidth]{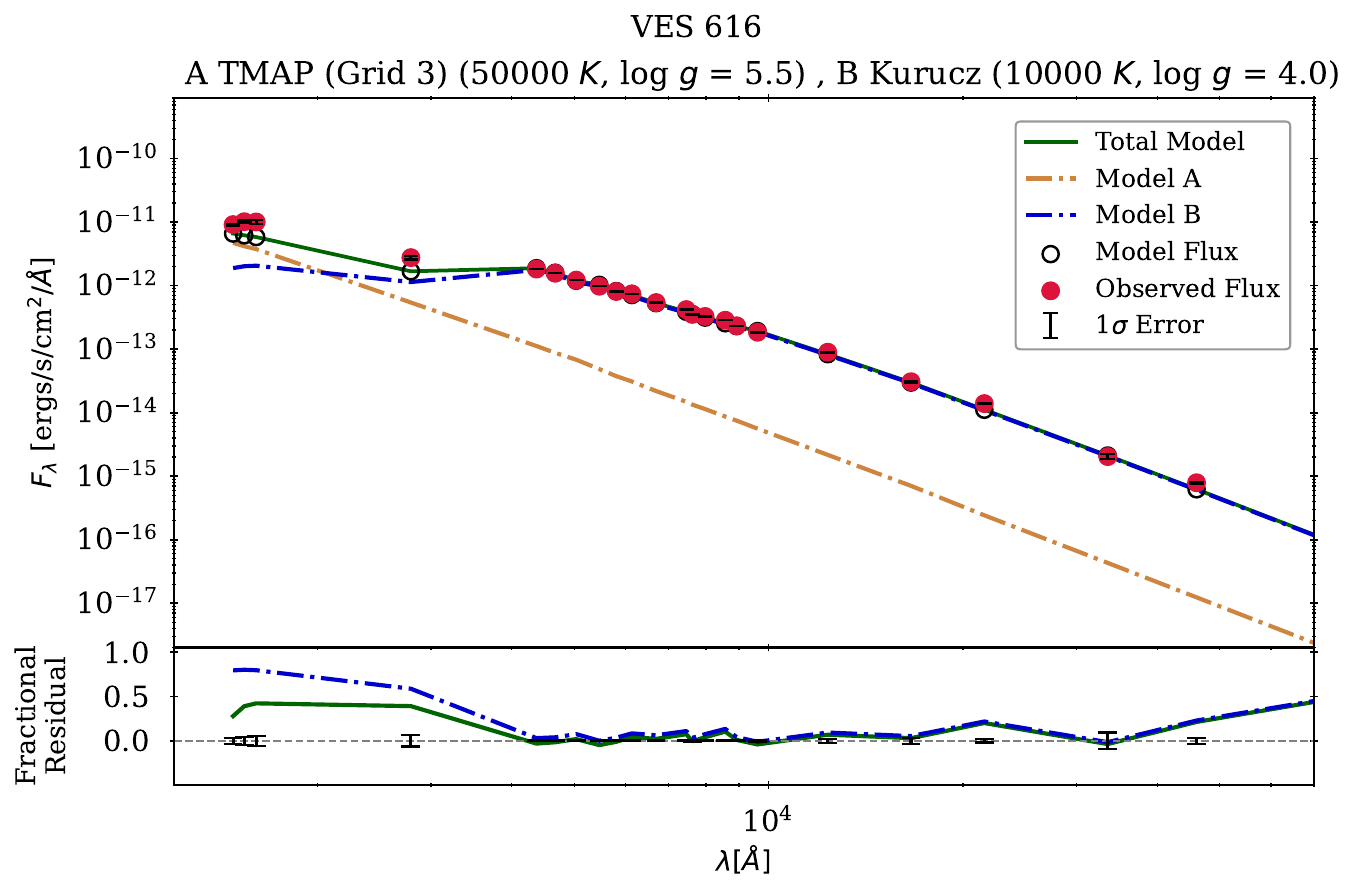}
    \includegraphics[width=0.41\columnwidth]{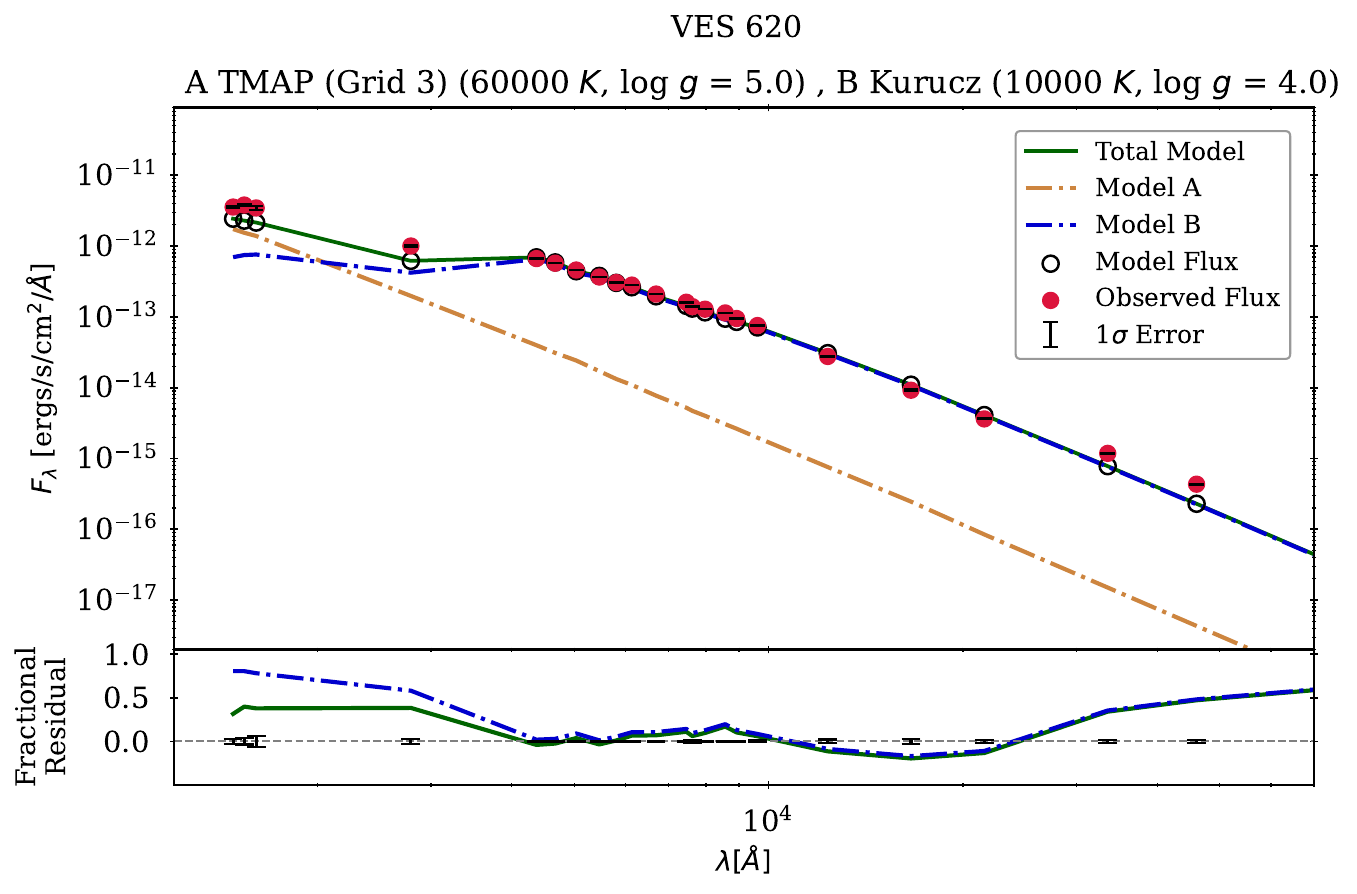}
    \includegraphics[width=0.41\columnwidth]{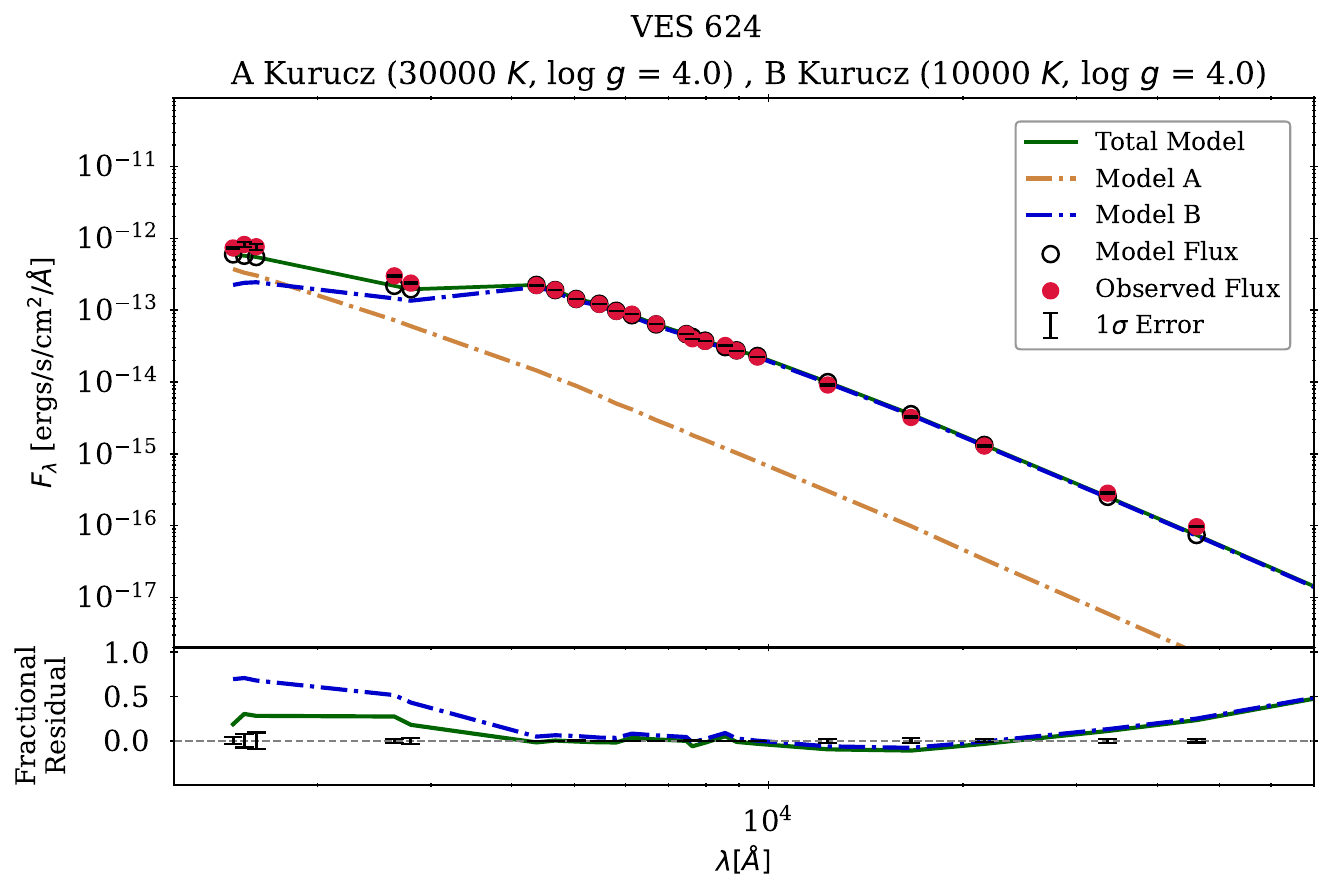}
    \caption{Continued.}
\end{figure*}

\begin{table*}
\small 
\setlength{\tabcolsep}{1.7pt} 
\renewcommand{\arraystretch}{1.1}
\centering
\caption{Fitting parameters for 23 detected Be stars in NGC~663.}
\label{comb_table}
\hspace*{-0.5cm}
\begin{tabular}{lllllllllllll}
\hline
Object & RA (Deg) & DEC (Deg) & P$_{PM}$ (\%) & Comp. & Model & T$_{eff}$ (K) & L/${L}_{\odot}$ & R/${R}_{\odot}$ & $\chi^2_{red}$ & Vgf & Vgf$_{b}$ & N$_{fit}$/N$_{tot}$ \\
\hline
\hline
\noalign{\vskip 2mm}
D01 034$^*$ & 26.765589 & 61.292236 & 68 & -- & Kurucz & 8750 $\pm$ 125 & 364 $\pm$ 69 & 8.34 $\pm$ 0.80 & 16.6 & 110 & 6.0 & 16/23 \\

GG 109$^*$ & 26.914015 & 61.305721 & 69 & -- & Kurucz & 9000 $\pm$ 125 & 366 $\pm$ 77 & 7.88 $\pm$ 0.75 & 21.4 & 5.8 & 0.3  & 16/23 \\

GG 95 & 26.443295 & 61.155804 & 61 & -- & Kurucz & 25000 $\pm$ 500 & 5255 $\pm$ 1075 & 3.96 $\pm$ 0.38 & 34.5 & 113 & 5.0  & 18/20 \\

L613 & 26.645269 & 61.107712 & 82 & -- & Kurucz & 10750 $\pm$ 125 & 176 $\pm$ 36 & 3.84 $\pm$ 0.37 & 18.2 & 59.9 & 6.7 & 19/24 \\

MWC 698 & 26.497110 & 61.212685 & 90 & -- & Kurucz & 14000 $\pm$ 500 & 5622 $\pm$ 1623 & 12.74 $\pm$ 1.21 & 3.9 & 53.6 & 3.9 & 17/21 \\

PKK 4 & 26.823006 & 61.221569 & 78 & -- & Kurucz & 11250 $\pm$ 125 & 91 $\pm$ 17 & 2.53 $\pm$ 0.24 & 30.3 & 20.8 & 1.5 & 13/16 \\

SAN 28$^*$ & 26.508623 & 61.250595 & 93 & -- & Kurucz & 10250 $\pm$ 125 & 123 $\pm$ 23 & 3.52 $\pm$ 0.34 & 16.9 & 7.6 & 0.5 & 18/21 \\

\noalign{\vskip 2mm}
\hline 
\noalign{\vskip 2mm}
BG 114 & 26.558376 & 61.228851 & 85 & Hot & Kurucz & 23000 $\pm$ 500 & 187 $\pm$ 35 & 0.85 $\pm$ 0.08 & 14.4 & 5.9 & 0.4 & 22/22 \\
 & & & & Cool & Kurucz & 11500 $\pm$ 125 & 313 $\pm$ 59 & 4.43 $\pm$ 0.42 & & & & \\

BG 15 & 26.615303 & 61.207062 & 100 & Hot & Kurucz & 42000 $\pm$ 500 & 4263 $\pm$ 812 & 1.23 $\pm$ 0.12 & 5.9 & 46.9 & 3.7 & 24/24 \\
 & & & & Cool & Kurucz & 11750 $\pm$ 125 & 3261 $\pm$ 621 & 13.70 $\pm$ 1.30 & & & & \\

G32 & 26.619238 & 61.230675 & 53 & Hot & TMAP (Grid 3) & 90000 $\pm$ 5000 & 1806 $\pm$ 344 & 0.17 $\pm$ 0.02 & 14.7 & 26 & 2.1 & 22/22 \\
 & & & & Cool & Kurucz & 11500 $\pm$ 125 & 404 $\pm$ 77 & 4.94 $\pm$ 0.47 & & & & \\

GC 97 & 26.483740 & 61.212574 & 82 & Hot & TMAP (Grid 3) & 50000 $\pm$ 5000 & 3881 $\pm$ 739 & 0.81 $\pm$ 0.08 & 4.4 & 47.4 & 5.5 & 24/24 \\
 & & & & Cool & Kurucz & 10000 $\pm$ 125 & 841 $\pm$ 160 & 9.47 $\pm$ 0.90 & & & & \\

GG 101 & 26.648338 & 61.227543 & 85 & Hot & Levenhagen & 58000 $\pm$ 500 & 573 $\pm$ 109 & 0.45 $\pm$ 0.04 & 16.0 & 129 & 11.2 & 24/24 \\
 & & & & Cool & Kurucz & 10000 $\pm$ 125 & 424 $\pm$ 81 & 6.69 $\pm$ 0.64 & & & & \\

GG 102 & 26.648041 & 61.263290 & 62 & Hot & Kurucz & 50000 $\pm$ 500 & 3616 $\pm$ 688 & 0.81 $\pm$ 0.08 & 1.7 & 5.3 & 0.9 & 20/20 \\
 & & & & Cool & Kurucz & 10750 $\pm$ 125 & 3035 $\pm$ 579 & 15.88 $\pm$ 1.51 & & & & \\

GG 108 & 26.861519 & 61.145599 & 53 & Hot & Kurucz & 36000 $\pm$ 500 & 1048 $\pm$ 199 & 0.80 $\pm$ 0.08 & 16.3 & 130 & 13.5 & 24/24 \\
 & & & & Cool & Kurucz & 11500 $\pm$ 125 & 1123 $\pm$ 214 & 8.23 $\pm$ 0.78 & & & & \\

GG 93 & 26.407560 & 61.133095 & 65 & Hot & Kurucz & 24000 $\pm$ 500 & 466 $\pm$ 89 & 1.19 $\pm$ 0.11 & 13.6 & 109 & 8.6 & 21/21 \\
 & & & & Cool & Kurucz & 10750 $\pm$ 125 & 532 $\pm$ 101 & 6.45 $\pm$ 0.61 & & & & \\

GG 94 & 26.415096 & 61.216434 & 68 & Hot & Kurucz & 39000 $\pm$ 500 & 704 $\pm$ 134 & 0.56 $\pm$ 0.05 & 8.9 & 43.9 & 5.3 & 24/24 \\
 & & & & Cool & Kurucz & 11500 $\pm$ 125 & 669 $\pm$ 127 & 6.35 $\pm$ 0.60 & & & & \\

MWC 700 & 26.612782 & 61.167202 & 82 & Hot & Kurucz & 39000 $\pm$ 500 & 4221 $\pm$ 804 & 1.41 $\pm$ 0.13 & 32.0 & 37.4 & 3.0 & 23/23 \\
 & & & & Cool & Kurucz & 12000 $\pm$ 125 & 3816 $\pm$ 727 & 14.15 $\pm$ 1.35 & & & & \\

PKK 2 & 26.449562 & 61.273327 & 62 & Hot & Kurucz & 35000 $\pm$ 500 & 148 $\pm$ 28 & 0.32 $\pm$ 0.03 & 7.1 & 64.3 & 4.3 & 24/24 \\
 & & & & Cool & Kurucz & 10500 $\pm$ 125 & 201 $\pm$ 38 & 4.23 $\pm$ 0.40 & & & & \\

PKK 3 & 26.601690 & 61.177017 & 95 & Hot & Kurucz & 25000 $\pm$ 500 & 70 $\pm$ 13 & 0.43 $\pm$ 0.04 & 13.1 & 30.4 & 5.5 & 24/24 \\
 & & & & Cool & Kurucz & 10250 $\pm$ 125 & 128 $\pm$ 24 & 3.53 $\pm$ 0.34 & & & & \\

VES 616 & 26.525537 & 61.227573 & 57 & Hot & TMAP (Grid 3) & 50000 $\pm$ 5000 & 2984 $\pm$ 568 & 0.70 $\pm$ 0.07 & 4.9 & 58.6 & 2.2 & 21/21 \\
 & & & & Cool & Kurucz & 12000 $\pm$ 125 & 664 $\pm$ 129 & 7.93 $\pm$ 0.75 & & & & \\

VES 619 & 26.611837  & 61.128258 & 91 & Hot & TMAP (Grid 3) & 50000 $\pm$ 5000 & 8095 $\pm$ 1542 & 1.18 $\pm$ 0.11 & 13.6 & 171 & 10.8 & 24/24 \\
 & & & & Cool & Kurucz & 10000 $\pm$ 125 & 1749 $\pm$ 337 & 13.80 $\pm$ 1.31 & & & & \\

VES 620 & 26.627619 & 61.241455 & 81 & Hot & TMAP (Grid 3) & 60000 $\pm$ 500 & 1768 $\pm$ 336 & 0.38 $\pm$ 0.04 & 9.8 & 124 & 10.7 & 24/24 \\
 & & & & Cool & Kurucz & 10000 $\pm$ 125 & 485 $\pm$ 92 & 7.06 $\pm$ 0.67 & & & & \\

VES 624 & 26.748318  & 61.208199 & 83 & Hot & Kurucz & 30000 $\pm$ 500 & 100 $\pm$ 19 & 0.35 $\pm$ 0.03 & 14.7 & 28.3 & 5.4 & 25/25 \\
 & & & & Cool & Kurucz & 10000 $\pm$ 125 & 151 $\pm$ 28 & 4.00 $\pm$ 0.38 & & & & \\

\hline
\end{tabular}
\tablefoot{
The upper portion lists stars fitted with a single-component model, while the lower portion includes stars fitted with a two-component model. Column 1 gives the object names used in this work. Columns 2 and 3 provide the RA and DEC of the stars. Column 4 lists the P$_{PM}$ of all Be stars in the cluster. Column 5 indicates the hot or cool component (in the case of a double fit), and column 6 specifies the models used. The T$_{eff}$, Ls, and Rs of the stars, along with their respective uncertainties, are presented in Columns 7, 8, and 9. Columns 10, 11, 12, and 13 list the $\chi^2_{\mathrm{red}}$, Vgf, and Vgf$_{b}$ value for the best fit and the ratio of the number of photometric data points used for the fit ($N_{\mathrm{fit}}/N_{\mathrm{tot}}$) to the total number of available data points, respectively. Also, objects marked with an asterisk ($^*$) show UV excess; these could not be fitted with a binary fit.}
\end{table*}

\end{appendix}

\end{document}